\documentclass[a4paper,11pt]{article}
\usepackage{jcappub} 
\usepackage[T1]{fontenc}
\usepackage{braket, xcolor, soul, cleveref,fullpage,tabularx, diagbox,hyperref}
\let\vec\mathbf
\renewcommand{\d}{\mathrm{d}}
\def\l{\left}
\def\r{\right}

\title{
Isocurvature Induced Gravitational Waves at Pulsar Timing Arrays}

\author[a,b]{Yi-Fu Cai}
\author[c]{Peizhi Du}
\author[a,b]{Jiahang Zhong}

\emailAdd{yifucai@ustc.edu.cn}
\emailAdd{dupeizhi@ustc.edu.cn}
\emailAdd{jiahangzhong@mail.ustc.edu.cn}

\affiliation[a]{Department of Astronomy, School of Physical Sciences, University of Science and Technology of China, Hefei 230026, China}
\affiliation[b]{CAS Key Laboratory for Research in Galaxies and Cosmology, School of Astronomy and Space Science, University of Science and Technology of China, Hefei 230026, China}
\affiliation[c]{Laboratory of Spin Magnetic Resonance, School of Physical Sciences, Anhui Province Key Laboratory of Scientific Instrument Development and Application, University of Science and Technology of China, Hefei 230026, China}

\abstract{Gravitational waves (GWs) are powerful probes of new physics in the early Universe. In particular, GWs induced by primordial isocurvature perturbations encode information of novel dynamics beyond the standard $\Lambda$CDM model. Existing studies of isocurvature induced GWs focus on a particular type: cold dark matter (CDM) isocurvature. In this work, we present a more comprehensive study of four kinds of isocurvature involving CDM, baryons, neutrinos and free-streaming dark radiation (DR). We first reformulate initial conditions of isocurvature with coupled neutrinos because modes relevant for observations at Pulsar Timing Arrays enter the horizon before neutrino decoupling. With these new initial conditions, neutrino isocurvature is phenomenologically similar to CDM isocurvature up to an overall coefficient, which leads to an interesting conversion of isocurvature between radiation and matter sectors.
We then find that the spectrum of isocurvature induced GWs from free-streaming DR is qualitatively different than that from CDM due to the presence of anisotropic stress. Unlike GWs induced by CDM isocurvature that are suppressed at high frequencies due to matter density being suppressed at early times, DR isocurvature induced GWs is proportional to the constant ratio between DR density and total radiation. 
Finally, we utilize two general parametrizations of the isocurvature power spectrum: a delta function and a broken power law, and derive novel constraints with recent NANOGrav data. Our results set stringent constraints on isocurvature around $10^{6}\,\textrm{Mpc}^{-1}$, which are complementary to cosmological observations at large scales.}

\begin{document}
\maketitle

\section{Introduction}\label{sec:introduction} 
The standard $\Lambda$CDM model can nicely explain current cosmological data across a wide range of scales~\cite{Planck:2018vyg}. In this model, the initial conditions are \textit{adiabatic}, meaning all metric and density perturbations are correlated. This is a natural consequence of standard inflation models where the primordial perturbations are dominantly sourced by fluctuations of a single scalar degree of freedom, typically the inflaton field. However, a large variety of physics beyond the Standard Model (BSM), which are proposed to address puzzles in the Standard Model (SM) and tensions in observations, can generate primordial perturbations via new dynamics independent of the inflaton fluctuations. This kind of initial conditions is called \textit{isocurvature}~\cite{Bucher:1999re,Wands:2000dp,Gordon:2000hv,Lyth:2002my,Malik:2004tf,Wands:2007bd,Doran:2003xq,Gerlach:2025hxw}. Isocurvature perturbations can leave distinct signatures in cosmological observations such as the cosmic microwave background (CMB), baryon acoustic oscillations (BAO), Lyman-$\alpha$ forest, and the Big Bang Nucleosynthesis (BBN)~\cite{Planck:2018jri,Planck:2019kim,Savelainen:2013iwa,Baumann:2015rya,Montandon:2020kuk,Kawakami:2012ke,Kawasaki:2011rc,Ghosh:2021axu,Adshead:2020htj,Dent:2012ne,Chluba:2013dna, Inomata:2018htm,Buckley:2025zgh,Gerlach:2025uxo,Chang:2025uvx,Kalia:2025uxg,Bagherian:2025puf,Gorghetto:2025uls}. Although current data have no definite preference for isocurvature, they provide novel constraints on new physics models and the potential to discover certain kinds of new physics with future measurements.

Gravitational waves (GWs) serve as another powerful probe of early-Universe physics. Because they free-stream, GWs can access information from much earlier times than the CMB photons. This prospect drives the continued, worldwide development of GW observatories designed to measure a broad range of frequencies. In particular, Pulsar Timing Arrays (PTA) provide a unique probe of GWs in the nano-Hertz frequency range.  Several experiments such as NANOGrav~\cite{NANOGrav:2023gor}, EPTA/InPTA~\cite{EPTA:2023fyk}, PPTA~\cite{Reardon:2023gzh}, CPTA~\cite{Xu:2023wog} have recently observed a stochastic GW background and derived constraints for new physics.  One possible source of GW that might contribute to PTA observations  is called \textit{scalar induced GWs} (SIGW)\footnote{See Ref.~\cite{Domenech:SIGWRe} for a review.}~\cite{Matarrese:1992rp,Matarrese:1993zf,Matarrese:1997ay,Ananda:2006af,Baumann:2007zm}: primordial scalar perturbations can source GWs at the second order in the perturbation theory. This framework has been applied to adiabatic perturbations and a certain type of isocurvature in cold dark matter (CDM)~\cite{Domenech:2021and,Domenech:2023jve,Chen:2024twp,Yuan:2024qfz,He:2024luf,Luo:2025lgr,Marriott-Best:2025sez,LaRosa:2025woi,Iovino:2025xkq,Han:2025wlo,Ali:2025xvj,Zeng:2025tno,Domenech:2025ffb,Papanikolaou:2025ddc,Garcia:2025yit,Yuan:2025seu}. In this work, we present a more comprehensive study of isocurvature induced GWs including four types of isocurvature: cold dark matter density isocurvature (CDI), baryon density isocurvature (BDI), neutrino density isocurvature (NDI) and free-streaming dark radiation density isocurvature (DRDI).  We emphasize that, for the frequency range relevant for PTA observations, the modes entered the horizon \textit{before} SM neutrino decoupling (when the temperature of the Universe is around MeV), and thus neutrinos are tightly coupled to the SM bath. Therefore, the initial conditions for NDI are qualitatively different from the standard ones where neutrinos are treated as free-streaming radiation. In this work, we derive novel initial conditions for all isocurvature modes with coupled neutrinos, and calculate the evolution of density and metric perturbations numerically.  We find that the GW spectrum from NDI with coupled neutrinos is the same as that of CDI or BDI up to an overall coefficient, which leads to an interesting `conversion' of isocurvature between radiation and matter sectors. Furthermore, DRDI results are dramatically different from those in the previously studied CDI case because of the presence of anisotropic stress. The size of DRDI induced GWs is significantly enhanced for high $k$ modes compared to that of CDI because the ratio of dark radiation (DR) density to the total bath is not suppressed at early times unlike the matter density. Therefore, PTA opens a new window to test BSM physics and places novel constraints via isocurvature induced GWs.

To derive general constraints that can be applied to a broad class of models, we use two parameterizations of the primordial isocurvature power spectrum: a delta function and a broken power law~\cite{Buckley:2025zgh}. Using the general formula for the source term of induced GWs, we obtain the spectrum of GWs and apply the data from NANOGrav collaboration to place constraints on the isocurvature power spectrum. Our results set stringent limits for wavenumbers around  $10^6\,\text{Mpc}^{-1}$, which are complementary to other cosmological constraints on large scales.

This paper is organized as follows. In section~\ref{sec:Iso_before_nu_dec} we present new isocurvature initial conditions before neutrino decoupling. We show a detailed calculation of GWs induced by general isocurvature perturbations in section~\ref{sec:IsoIGWs}. We derive constraints on the isocurvature power spectrum from PTA in section~\ref{sec:PTA}. We conclude in section~\ref{sec:conclusions}.

\section{Isocurvature Initial Conditions Before Neutrino Decoupling}
\label{sec:Iso_before_nu_dec}
The standard isocurvature initial conditions are defined based on the $\Lambda$CDM model relevant to the physics around the era of CMB formation (see e.g., Ref.~\cite{Bucher:1999re}). In particular, neutrinos are treated as free-streaming radiation since they have decoupled from the SM bath a long time before recombination. However, the modes that can source GWs accessible at PTA enter the horizon when the temperature of the universe is around $100$ MeV scale and neutrinos are tightly coupled to the SM bath. Moreover, the radiation bath around that period consists of not only photons and neutrinos, but also electrons and positrons, or even muons and pions. Therefore, a new formulation of isocurvature perturbations is required. 

To study perturbations in cosmology~\cite{Kodama:1986fg,Kodama:1986ud,Ma:1995ey}, we start with the perturbed FRW metric in the conformal Newtonian gauge,
\begin{eqnarray}\label{eq:perturbed_metric}
    d s^2 = a^2(\tau)\left[-(1+2\Phi)d \tau^2+(\delta_{ij}-2\Psi \delta_{ij}+h_{ij}) d x^i d x^j\right]~,
\end{eqnarray}
where $a$ is the scale factor and $\tau$ is the conformal time. $\Phi$ and $\Psi$ are two scalar metric perturbations. $h_{ij}$ denotes the tensor perturbation that satisfies the transverse and traceless conditions: $\partial^i h_{ij}=0, h^{i}_i=0$. 

We consider the early Universe is dominated by radiation and matter, and thus the total background energy density $\bar\rho$ is given as 
\begin{eqnarray}
\bar\rho&=&\bar\rho_{r}+\bar\rho_m.
\end{eqnarray}
Here $\bar\rho_{r/m}$ is the background energy density of radiation and matter, which satisfies the following evolution equations:
\begin{eqnarray}
    \bar\rho_{r}'+4\mathcal H \bar\rho_r=0~~;~~ \bar\rho_{m}'+3\mathcal H \bar\rho_m=0,
\end{eqnarray}
where the prime means the derivative with respect to the conformal time $\tau$. $\mathcal H\equiv a'/a$ is the conformal Hubble parameter which can be written as
\begin{eqnarray}
    \mathcal H(\tau)=\frac{1}{\tau}\frac{1+\omega\tau/2}{1+\omega\tau/4},
\end{eqnarray}
where $\omega\equiv \Omega_{m} H_0/\sqrt{\Omega_r}$ with $H_0$ being the Hubble rate today and $\Omega_{m/r}\equiv \bar\rho_{m/r,0}/\rho_{c,0}$ is the fractional energy density of matter/radiation today with respect to the present-day critical density $\rho_{c,0}$.

The radiation bath contains photons ($\gamma$), neutrinos ($\nu$), and may also have other light particles in the SM as well as DR.  
To simplify the discussion, we treat the \textit{adiabatic} SM radiation bath as one species (denoted as $\gamma_*$), which contains all SM radiation components that do not have relative entropy perturbations (defined in Eq.~\eqref{eq:S_ab}). This approach is legitimate because all these components have the same initial conditions and evolution functions.  We note that for most cases $\nu$ is a part of $\gamma_*$, but we separate it from $\gamma_*$ for the NDI case since it contains isocurvature.

Now we move on to study different types of perturbation in cosmology. The gauge invariant curvature perturbation can be written in the Newtonian gauge as~\cite{Wands:2000dp}
\begin{align}
    \zeta = -\Psi -{\cal H}\frac{\delta \rho}{\rho'}~.
    \label{eq:curvature}
\end{align}
where $\delta\rho=\rho-\bar\rho$ denotes the density perturbation.  One may also define a curvature fluctuation for each species, namely
\begin{align}
    \zeta_\alpha = -\Psi-{\cal H} \frac{\delta \rho_\alpha}{\rho'_\alpha}~.
\end{align}
where $\alpha\in\{\gamma_*,\nu,b,c,\rm dr\}$ labels the species in the Universe. We further define the perturbation of the total radiation bath as
\begin{eqnarray}
    \zeta_{r}\equiv\sum_{i\in\,r} R_{i} \zeta_{i},
\end{eqnarray}
where $R_i\equiv \bar\rho_{i}/\bar\rho_r$ with $\bar\rho_r$ being the total background radiation density and $i$ denotes species in the radiation bath.
The gauge-invariant isocurvature perturbation, or the relative entropy perturbation is defined as
\begin{eqnarray}\label{eq:S_ab}
    {\cal S}_{\alpha\beta}  = 3(\zeta_\alpha-\zeta_\beta)=\frac{\delta_\alpha}{1+w_\alpha}-\frac{\delta_\beta}{1+w_\beta}~,
\end{eqnarray}
where $w\equiv\bar p/\bar \rho$ with $\bar p$ being the background momentum density.

Now we are ready to define initial conditions for all metric and density perturbations in the superhorizon limit $k\tau\ll 1$, where $k$ is the wavenumber of the perturbation. Given the evolution functions from the Einstein equations and Boltzmann equations in the superhorizon limit, there are sets of linearly independent solutions of all perturbations. These solutions are named adiabatic and isocurvature initial conditions. We can then write perturbation variables in each solution as
\begin{eqnarray}\label{eq:split_modes}
    X(\vec{k},\tau)=c(\vec{k}) X(k,\tau),
\end{eqnarray}
where $X\in\{\Psi,\Phi,\delta_\alpha,\theta_\alpha,\sigma_\alpha\}$ with $\delta\equiv\delta\rho/\bar\rho,\theta,\sigma$ being the density, velocity perturbations, and anisotropic stress respectively. Here $\vec k \equiv k \hat{\vec{k}}$ is the Fourier conjugate to coordinate $\vec x$. $c(\vec{k})$ is a time-independent coefficient which is common for all perturbation variables in a certain set of initial conditions (but different for different initial conditions), which encodes the primordial spacial distribution of perturbations. $X(k, \tau)$ is perturbation-specific but independent of the direction $\vec{\hat{k}}$, which is also called the transfer function. For the rest of this section, we show initial conditions for $X(k,\tau)$.

One of the solutions encodes non-vanishing curvature perturbation ($\zeta\neq0$) with no relative entropy perturbations ($\mathcal S_{\alpha\beta}=0$) in the limit $k\tau\to 0$, which is called adiabatic (AD) initial conditions. Fixing $\zeta=1$ and keeping the leading term as $k\tau\to0$, the adiabatic initial conditions are given by
\begin{eqnarray}
       \Psi&=&\Phi=-\frac{2}{3}~ \nonumber\\
    \delta_{c}&=&\delta_{b}=\frac{3}{4}\delta_{\gamma_*}=1~~~~~~~~~~~~~~~~~(\textrm{AD})\nonumber\\
    \theta_{c}&=&\theta_{b}=\theta_{\gamma_*}=-\frac{1}{3}k^2\tau. \label{eq:AD}
\end{eqnarray}

All other solutions are called isocurvature initial conditions, which have vanishing curvature perturbation $\zeta=0$ but at least one non-zero $\cal S_{\alpha\beta}$. Since there are multiple species in the Universe, we can define different isocurvature initial conditions by different choices of non-vanishing $\cal S_{\alpha\beta}$. In this paper,  we choose the adiabatic SM radiation bath  $\gamma_*$ as the reference and we consider only one non-vanishing $\mathcal S_{\alpha\gamma_*}$ in each case as $\tau\to0$.\footnote{Since $\mathcal S_{\alpha\gamma_*}$ is always the same as $\mathcal S_{\alpha\gamma}$, it is equivalent to choose the photon bath as the reference.} For example, NDI is defined by setting  $\mathcal S_{\alpha\gamma_*}=0$ except for $\mathcal S_{\nu\gamma_*}\neq 0$. Similarly, we can define BDI ($\mathcal S_{b\gamma_*}\neq 0$), CDI ($\mathcal S_{c\gamma_*}\neq 0$), and DRDI ($\mathcal S_{\textrm{dr}\gamma_*}\neq 0$). {Once fixing the value of the non-zero $\mathcal S_{\alpha\gamma_*}$ (e.g., we choose $\mathcal S_{c\gamma_*}=1$ for CDI), initial conditions for all metric and density perturbations are uniquely determined.} We note that the definition of isocurvature initial conditions may vary  depending on the choice of non-vanishing $\cal S_{\alpha\beta}$. Another commonly studied choice is based on $\mathcal S_{\alpha r}$ which involves the perturbation of the radiation bath~\cite{Bucher:1999re}. We find this choice matches our definition if the isocurvature component is not in the radiation bath like BDI and CDI, but differs for NDI and DRDI. 
Since $\zeta_{r}$ in the NDI case contains neutrino perturbations that have isocurvature, setting $\mathcal S_{b r/cr}= 0$ does not respect the adiabaticity of baryons and CDM. Therefore, for the study of new physics that generate isocurvature in certain species and the rest remain adiabatic, it is more convenient to use the definition based on $\mathcal S_{\alpha\gamma_*}$ instead of $\mathcal S_{\alpha r}$.

With this definition and setting $\mathcal S_{b\gamma_*/c\gamma_*}=1$, metric and density perturbations in BDI/CDI can be written as (keeping to the order $\omega\tau$)
\begin{eqnarray}
     \Psi&=&\Phi=-\frac{1}{8}\omega_{b}\tau~ \nonumber\\
    \delta_c&=&\delta_b-1=\frac{3}{4}\delta_{\gamma_*}=- \frac{3}{8}\omega_{b}\tau ~~~~~~~~~~~~~~~(\textrm{BDI})\nonumber\\
    \theta_{c}&=&\frac{1}{3}\theta_{b}=\frac{1}{3}\theta_{\gamma_*}=-\frac{1}{24}\omega_{b}k^2\tau^2  \label{eq:BDI}\\
    \Psi&=&\Phi=-\frac{1}{8}\omega_{c}\tau~ \nonumber\\
    \delta_b&=&\delta_c-1=\frac{3}{4}\delta_{\gamma_*}=- \frac{3}{8}\omega_{c}\tau ~~~~~~~~~~~~~~~(\textrm{CDI})\nonumber\\
    \theta_{c}&=&\frac{1}{3}\theta_{b}=\frac{1}{3}\theta_{\gamma_*}=-\frac{1}{24}\omega_{c}k^2\tau^2, \label{eq:CDI}
\end{eqnarray}
where $\omega_{b/c}\equiv (\Omega_{b/c}/\Omega_m)\omega$ and $\Omega_{b/c/m}$ is the fractional energy density of baryons/CDM/matter today. 

As mentioned before, we separate $\nu$ from $\gamma_*$ for NDI. Setting $\mathcal S_{\nu\gamma_*}=3/4$,\footnote{This is the convention commonly used for radiation isocurvature, which ensures the leading isocurvature perturbation (e.g., $\delta_\nu$ in NDI) is normalized close to unity~\cite{Bucher:1999re}. } we get perturbations in NDI as 
\begin{eqnarray}
     \Psi&=&\Phi=\frac{3}{32}R_\nu\omega\tau~ \nonumber\\
    \delta_{\gamma_*}&=&\delta_\nu-1=\frac{4}{3}\delta_{b/c}=- R_\nu+\frac{3}{8}R_\nu\omega\tau~~~~~~~~~~~~~~~(\textrm{NDI})\nonumber\\
    \theta_{c}&=&\frac{1}{3}\theta_{b}=\frac{1}{3}\theta_{\gamma_*}=\frac{1}{3}\theta_{\nu}=-\frac{1}{24}R_\nu\omega k^2\tau^2. \label{eq:NDI}
\end{eqnarray}
{Although $\delta_\alpha$ approaches a constant as $\tau\to 0$, they do not source metric or curvature perturbations because the total density perturbation $\delta=(\sum_\alpha \bar\rho_\alpha\delta_\alpha)/\bar\rho$ vanishes as $\tau\to 0$.}

Now we are ready to study the interesting relation between NDI and BDI/CDI. To clearly see the relation, we treat the radiation bath as a whole and its perturbation is defined as $\delta_r\equiv(1-R_\nu)\delta_{\gamma_*}+R_\nu \delta_\nu$: $\delta_r=\delta_{\gamma_*}$ (BDI/CDI) and $\delta_r=(3/8)R_\nu\omega\tau$ (NDI). Then it is obvious that NDI is simply a rescaling of BDI+CDI:
\begin{eqnarray}
    X^{\rm NDI}= -\frac{3}{4}R_\nu( X^{\rm BDI}+X^{\rm CDI}).
\end{eqnarray}
This can be understood that NDI generates net entropy perturbations between radiation and matter sectors, which effectively act as matter isocurvature with the opposite sign. Since the evolution of BDI and CDI perturbations is almost the same with a simple rescaling, the observational signature of NDI is basically the same as BDI or CDI up to an overall coefficient, which exhibits an interesting ``conversion'' between NDI and matter isocurvature. 

Additional free-streaming DR is introduced for DRDI. Free-streaming DR develops a sizable anisotropic stress, which sources the difference between $\Psi$ and $\Phi$ via the Einstein equations (in radiation domination)
\begin{eqnarray}\label{eq:Psi_Phi_diff}
    \Psi-\Phi = 6(\mathcal H/k)^2 \sigma,
\end{eqnarray}
where $\sigma$ denotes the total anisotropic stress. Setting $\mathcal S_{{\rm dr}\gamma_*}=3/4$, perturbations in DRDI are given as (keeping the leading order in $k\tau$)
\begin{eqnarray}\label{eq:transfer_DRDI}
      \Psi&=&-\frac{1}{2}\Phi=\frac{(1-R_{\rm dr})R_{\rm dr}}{15+4R_{\rm dr}}~ \nonumber\\
    \delta_{\gamma_*}&=&\delta_{\rm dr}-1=\frac{4}{3}\delta_{b/c}=-\frac{R_{\rm dr}(11+8R_{\rm dr})}{15+4R_{\rm dr}}\nonumber\\
    \theta_{c}&=&-\frac{(1-R_{\rm dr})R_{\rm dr}}{15+4R_{\rm dr}}k^2\tau\nonumber\\
    \theta_{b}&=&\theta_{\gamma_*}=-\frac{19R_{\rm dr}}{60+16R_{\rm dr}}k^2\tau\nonumber\\
    \theta_{\rm dr}&=&\frac{15(1-R_{\rm dr})}{60+16R_{\rm dr}}k^2\tau~~~~~~~~~~~~~~~~~~~~~~~~~~~~~~~~~~(\textrm{DRDI})\nonumber\\
    \sigma_{\rm dr}&=&\frac{(1-R_{\rm dr})}{2(15+4R_{\rm dr})}k^2\tau^2. \label{eq:DRDI}
\end{eqnarray}
Unlike BDI/CDI or NDI, DRDI has non-vanishing $\Psi$ and $\Phi$ as $\tau\to0$. This means $\Psi$ and $\Phi$ can be sizable at superhorizon scales, although they are proportional to $R_{\rm dr}$ when $R_{\rm dr}\ll 1$. This leads to qualitatively new features in induced GWs, as we will discuss in the next section.

\begin{figure}[!tb]
\centering
\includegraphics[width=0.48\textwidth]{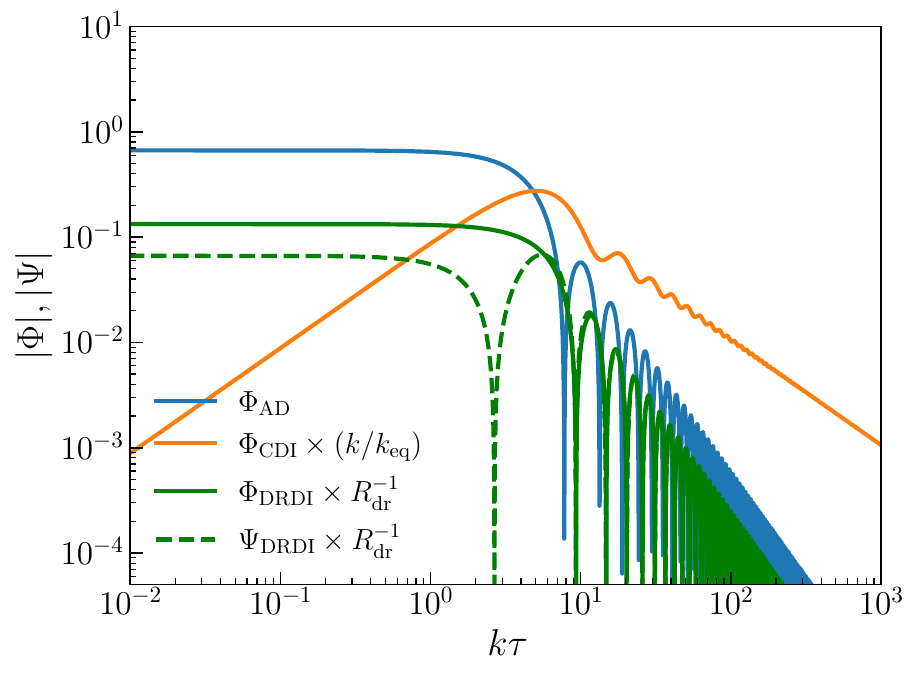}
\includegraphics[width=0.48\textwidth]{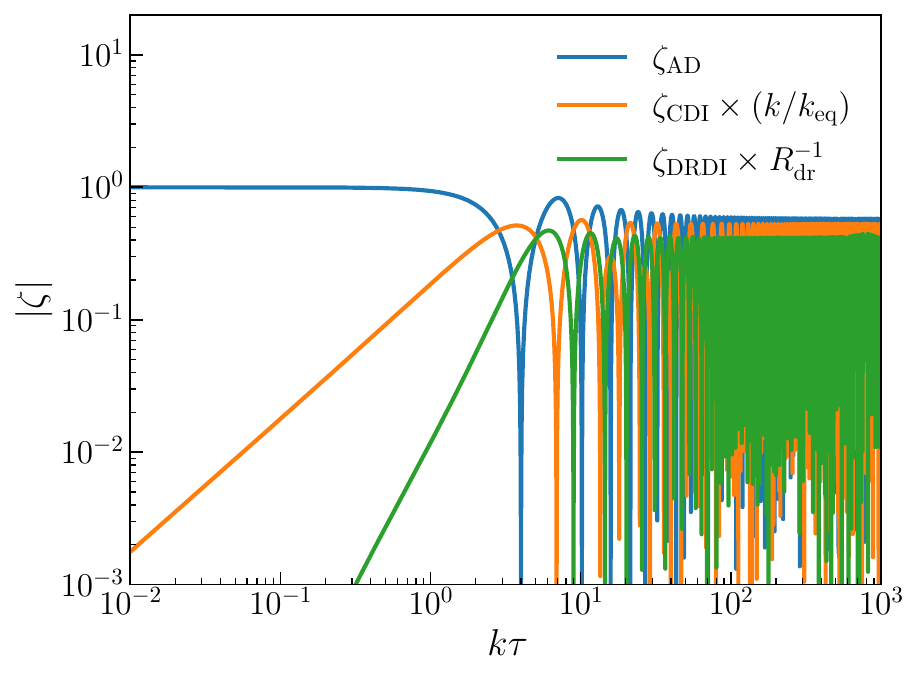}
\caption{The absolute value of transfer functions of metric perturbations $|\Phi|$ and $|\Psi|$ in the Newtonian gauge (left) and curvature perturbation $|\zeta|$ (right) as a function of $k\tau$ for AD, CDI and DRDI. $\Phi=\Psi$ for AD and CDI due to no anisotropic stress. $k_{\rm eq}$ is the wavenumber of the mode that enters the horizon at the matter radiation equality, and $R_{\rm dr}\equiv \bar\rho_{\rm dr}/\bar\rho_{r}$ is the fraction of the DR density to the total radiation density. The scaling of the CDI (DRDI) case is valid for $k\gg k_{\rm eq}$ ($R_{\rm dr}\ll1$). }
\label{fig:transfer}
\end{figure}

We show the time evolution of $\Psi$ and $\Phi$ for different cases in the left panel of figure~\ref{fig:transfer}. For illustrative purposes,  we just show the AD, CDI and DRDI cases. Since initial conditions for CDI is qualitatively similar to BDI and NDI up to an overall rescaling as mentioned before, these choices covers all relevant features.
As seen from figure~\ref{fig:transfer}, $\Psi=\Phi\propto C$ for AD and $\Psi=\Phi\propto \omega\tau$ for CDI at superhorizon. They both become damped oscillations after entering the horizon. Around horizon crossing ($k\tau\sim1$), the CDI case is suppressed by $\omega/k$ (or $k_{\rm eq}/k$ because $k_{\rm eq}=\sqrt{2}\omega$) for large $k$. Therefore we rescale the CDI case by the inverse of this factor. The DRDI case, however, has qualitatively different $\Phi$ and $\Psi$ evolutions due to the presence of anisotropic stress from free-streaming DR. In DRDI, $\Phi,\Psi,\Phi-\Psi\propto R_{\rm dr}$ (see Eq.~\eqref{eq:DRDI}) at superhorizon. They both start damped oscillations after horizon entry and become the same deep inside the horizon due to the  anisotropic stress being suppressed (see Eq.~\eqref{eq:Psi_Phi_diff}). Moreover, we present the evolution of the curvature perturbation $\zeta$ in the right panel of figure~\ref{fig:transfer}.  $\zeta$ of AD starts at one by definition and oscillates after horizon entry. For isocurvature cases, $\zeta$ starts at zero but grows as time evolves and oscillates after horizon entry. We note that for cases without anisotropic stress (AD and CDI), $\zeta$ behaves qualitatively similarly to $\Phi$ or $\Psi$. However, $\zeta$ of DRDI is dramatically different than $\Phi$ or $\Psi$ due to the net anisotropic stress.

\paragraph{Primordial Power Spectrum}
Another important information from initial conditions is the initial spatial distribution encoded in $c(\vec k)$ defined in Eq.~\eqref{eq:split_modes}. Given the random nature of the initial distribution, we often study observables in terms of this two-point function
 \begin{eqnarray}\label{eq:primordial_P}
   \langle c(\vec{k})c(\vec{k'}) \rangle  \equiv \frac{2\pi^2}{k^3}\mathcal P(k)(2\pi)^3\delta^3(\vec k +\vec{k}'),
 \end{eqnarray}
 where $\mathcal P(k)$ is the dimensionless power spectrum. 
For different initial conditions studied in this paper, we get
 \begin{eqnarray}
     \mathcal P(k) =\left\{
     \begin{array}{cc}
         \mathcal P_\zeta(k) &(\textrm{AD})  \\
         \mathcal P_{\mathcal S}(k)&(\textrm{BDI/CDI}) \\ 
         \frac{16}{9}\mathcal P_{\mathcal S}(k)&(\textrm{NDI/DRDI})
     \end{array}
     \right.,
 \end{eqnarray}
where $\mathcal P_\zeta$ and $\mathcal P_{\mathcal S}$ are the dimensionless power spectra for $\langle\zeta\zeta\rangle$ and $\langle\mathcal S \mathcal{S}\rangle$, respectively. Notice that the factor $16/9$ arises because transfer functions for NDI/DRDI modes are defined by setting ${\cal S}=3/4$ as $\tau\to0$ (see discussions above).

To obtain general constraints on isocurvature from various new physics models, we parameterize $\mathcal P(k)$ as the following two general forms: a delta function and a broken power law. The delta function spectrum has the form
\begin{eqnarray}\label{eq:P_delta}
    \mathcal P_{\rm iso}(k)=A_{\rm iso}\,\delta(\ln k -\ln k_0),
\end{eqnarray}
where $A_{\rm iso}$ is the amplitude and $k_0$ is the wavenumber where the delta function peaks. While no physical models provide the exact delta function power spectrum, it allows us to isolate the effect of
isocurvature on relevant observables at each wavenumber. Therefore, this constraint can be applied to any extended spectrum up to $O(1)$ factors assuming no correlation among different wavenumbers. Another more realistic parameterization is a broken power law spectrum given as
\begin{eqnarray}\label{eq:P_Broken_Powerlaw}
    {\cal P}_{\rm iso}(k) = A_{\rm iso}\left\{\begin{matrix} 
  (k/k_0)^3 & k\leq k_0 \\  
  1 & k>k_0
\end{matrix}\right.~,
\end{eqnarray}
where $k_0$ denotes the scale where the transition occurs. The $k^3$ dependence for small $k$ is generic for all modes that are outside the horizon around the generation of isocurvature (also known as white noise spectrum, see e.g., \cite{Graham:2015rva,Irsic:2019iff,Feix:2020txt,Amin:2022nlh,Redi:2022llj,Elor:2023xbz,Graham:2024hah,deKruijf:2024voc,Ling:2024qfv,Amin:2025dtd}). If the observation data are mostly from this $k^3$ tail (i.e., for large $k_0$), the constraint can be applied to any model that has a $k^3$ spectrum up to a rescaling. The $k^0$ part for $k>k_0$ can originate from subhorizon dynamics during inflation, which naturally has nearly scale invariance. One concrete model that has both features is the non-thermal phase transition during inflation~\cite{Buckley:2024nen,Barir:2022kzo}. In this case, we can map the parameters in $\mathcal P(k)$ to model parameters. The amplitude $A_{\rm iso}\sim \Gamma_{\rm PT}/H_{\rm inf}^4$, where $\Gamma_{\rm PT}$ is the phase transition rate per volume and $H_{\rm inf}$ is the Hubble rate during inflation. The comoving scale when this phase transition starts is related to $k_0$.

We also show results for the log-normal power spectrum, which is widely studied for induced GWs~\cite{Pi:2020otn}. The log-normal spectrum is given as
\begin{eqnarray}\label{eq:P_lognormal}
     {\cal P}_{\rm iso}= \frac{A_{\rm iso}}{\sqrt{2\pi}\Delta} \exp\left(-\frac{\ln^2(k/k_0)}{2\Delta^2}\right)~,
\end{eqnarray}
where $\Delta$ denotes the dimensionless width. This spectrum can be thought as a smeared version of the delta function spectrum, which can arise in certain inflationary models~\cite{Pi:2017gih,Kawasaki:1997ju,Frampton:2010sw,Cai:2018tuh}. 
\paragraph{Particle Physics Models for Isocurvature Perturbations}
To source isocurvature perturbations, it generally needs new sources of perturbations other than the inflaton fluctuations. Additional scalar fields during inflation, such as axion or curvaton models (for reviews, see e.g.,~\cite{Marsh:2015xka,Mazumdar:2010sa}), can source isocurvature after they decay. Other new dynamics like phase transitions during inflation can also excite isocurvature after they convert the vacuum energy of the phase transition into other species. If the decay or energy transfer is to the decoupled sector, such as CDM and DR, the isocurvature is straightforwardly inherited from the perturbations of the source. However, the BDI and NDI cases are subtler since baryons and neutrinos are tightly coupled to the SM bath and can efficiently transfer energy among them. The key point to maintain net isocurvature in baryons (neutrinos) is that the source must violate baryon (lepton) number, and there are no baryon or lepton number violation interactions in the thermal bath. Therefore the baryon (lepton) number is conserved and maintain the net isocurvature even they are in the thermal equilibrium. Such conditions can be achieved via a late decay of a particle that violates baryon or lepton number after the Electro-Weak sphaleron going out of the bath (when $T\sim 100$ GeV). For example, considering the thermal bath is initially adiabatic with no chemical potential in baryons (leptons), the number violating decays generate a net chemical potential $\mu$ in the phase space distribution of baryons (leptons) and opposite for antibaryons (antileptons). The total energy density of baryons and antibaryons differs from the case without chemical potential (adiabatic case) at $O(\mu^2)$ for $\mu\ll T$, which becomes the isocurvature. Therefore, in these cases, the amplitude of isocurvature power spectrum is related to the second order of the chemical potential.

\paragraph{Comparing Isocurvature Initial Conditions Before and After Neutrino Decoupling}
For comparison, we list the results for conventional definition (with free-steaming neutrinos) of various isocurvature modes in appendix~\ref{app:Iso_after_nu_dec}, but with $\mathcal S_{\alpha\gamma_*}\neq0$ instead of $S_{\alpha r}\neq0$. Here we summarize the main differences between isocurvature modes before and after neutrino decoupling. 
\begin{itemize}
    \item The main difference comes from the fact that neutrinos are coupled to the SM bath before decoupling, but become free-steaming after decoupling. This leads to vanishing anisotropic stress $\sigma$ for most cases before decoupling, except for DRDI. Therefore, $\Psi=\Phi$ is valid except for DRDI (see Eq.~\eqref{eq:Psi_Phi_diff}). Whereas modes after decoupling always have anisotropic stress from free-steaming neutrinos.
    \item For AD, BDI and CDI modes, initial conditions before neutrino decoupling can be simply obtained from ones after neutrino decoupling by setting $R_\nu=0$ and removing perturbations from neutrinos.
    \item As mentioned before, NDI is qualitatively different before and after neutrino decoupling. Remarkably, NDI is phenomenologically similar to CDI or BDI  before decoupling, which reflects an interesting ``conversion'' between radiation and matter isocurvature. {Transition between NDI before and after neutrino decoupling is non-trivial, which requires to study the full Boltzmann equations with proper time-dependent collision terms~(see e.g.,~Refs.~\cite{Oldengott:2014qra,Oldengott:2017fhy}). Detailed study of this transition is beyond the scope of current work.}
    \item Since free-streaming DR behaves like free-streaming neutrinos, it is equivalent between DRDI before neutrino decoupling and the conventional NDI by exchanging $R_{\rm dr}$  with $R_\nu$.
    
\end{itemize}

\section{Isocurvature Induced Gravitational Waves}\label{sec:IsoIGWs}
With just scalar (metric and density) perturbations, tensor fluctuations $h_{ij}$ in Eq.~\eqref{eq:perturbed_metric} can not be sourced at the first order in perturbation theory due to the mismatch of the spin structure. However, at the second order, there is a source term for tensor modes from the combination of scalar perturbations. 
{The equation of motion for tensor fluctuation $h_{ij}$ can be derived from the second-order Einstein's equation, which depends on the second-order energy momentum tensor $T^{(2)}_{ij}$. For cases without anisotropic stress (AD/CDI/BDI/NDI), the whole radiation bath is strongly coupled. Under the tight-coupling approximation, the leading contribution to the second-order energy momentum tensor is given as~\cite{Saga:2014jca,Mangilli:2008bw}}
\begin{eqnarray}
    T^{(2)}_{ij}=\frac{4}{3}\bar\rho_{r}\partial_i v_{\gamma_*}\partial_jv_{\gamma_* }+(\textrm{diagonal part})\delta_{ij}\,~~~(\textrm{without DR}),
\end{eqnarray}
{where $v$ is another way to denote the velocity perturbation, which is related to the previously defined $\theta$ as $\theta(k)=-k^2v(k)$.\footnote{We drop the superscript label $^{(1)}$ for first order perturbations for simplicity.} Here we neglect contributions from matter components (CDM and baryons) because they are highly suppressed deep inside the radiation domination. For the DRDI case, free-streaming DR cannot be treated as a coupled fluid, and thus the second-order energy momentum tensor reads}
\begin{eqnarray}\label{eq:T_dr_(2)}
    T^{(2)}_{ij}=\frac{4}{3}\bar\rho_{\gamma_*}\partial_iv_{\gamma_*}\partial_jv_{\gamma_*}+\bar\rho_{\textrm{dr}}\,\Pi^{(2)}_{\textrm{dr}\,ij}+(\textrm{diagonal part})\delta_{ij}\,~~~(\textrm{DRDI}),
\end{eqnarray}
{where $\Pi^{(2)}_{\textrm{dr}\,ij}$ denotes the contribution of DR to the second-order energy momentum tensor. $\Pi^{(2)}_{\textrm{dr}\,ij}$ is determined by the Boltzmann equation at the second order, which is not simply $\propto \partial_iv_{\textrm{dr}}\partial_j v_{\textrm{dr}}$ (see Refs.~\cite{Saga:2014jca,Mangilli:2008bw, Weinberg:2003ur, Zhang:2022dgx} and appendix~\ref{app:Boltzmann_DR}).}

Following the standard formalism in Refs.~\cite{Kohri2018,Espinosa:2018eve,Pi:2020otn,Gong:2019mui}, the equation of motion for tensor perturbation $h_{ij}$ (or gravitational waves) is given by
\begin{eqnarray}\label{eq:eom_h_x}
    h''_{ij}+ 2 {\cal H} h'_{ij}-\nabla^2 h_{ij} = S_{ij}~,
\end{eqnarray}
where the source term for AD/CDI/BDI/NDI cases in the radiation dominated Universe is~\cite{Domenech:SIGWRe}
\begin{align}\label{eq:source_woDR}
    S_{ij} =& 4{\cal P}^{ab}{}_{ij}\left[\partial_a\Phi\partial_b\Phi+
    4\pi G a^2 \frac{4}{3}\bar \rho_r \partial_a v_{\gamma_*}\partial_bv_{\gamma_*}\right]~~~~(\textrm{without DR}),
\end{align}
where \({\cal P}^{ab}{}_{ij}\) is a transverse and traceless operator and we used $\Psi=\Phi$ due to the absence of anisotropic stress. We note that NDI has the same expression because $\bar\rho_r=\bar\rho_{\gamma_*}+\bar\rho_{\nu}$ and $v_{\gamma_*}=v_{\nu}$.

After the Fourier transform, the equation of motion becomes
\begin{eqnarray}\label{eq:diff_h}
    h''_{\vec{k},\lambda}+ 2 {\cal H} h'_{\vec{k},\lambda}+k^2h_{\vec{k},\lambda} = S_{\vec{k},\lambda}~
\end{eqnarray}
with \(\lambda=+,\times\) denoting two polarizations of $h_{ij}$ and 
\begin{eqnarray}\label{eq:general_source}
    S_{\vec{k},\lambda}=4\int \frac{d^3 \vec{q}}{(2\pi)^3}e^{ij}_{\vec{k}, \lambda}q_iq_j {c (\vec q)} c(\vec{k}-\vec{q})f(\tau,q,|\vec{k}-\vec{q}|)~,
\end{eqnarray}
where $e^{ij}_{\vec k, \lambda}$ is the polarization tensor and
\begin{eqnarray}\label{eq:f_function_woDR}
    f(\tau,q,|\vec k- \vec q|) &=& \Phi(q,\tau)\Phi(|\vec k- \vec q|,\tau)~~~~~~~~~~~~~~~~~~~~~~~~~~~~~~~~~~~~~(\textrm{without DR})\\
    &~&+\frac{1}{2}\left(\Phi(q,\tau)+\frac{\partial_\tau \Phi(q,\tau)}{\cal H}\right)\left(\Phi(|\vec k- \vec q|,\tau)+\frac{\partial_\tau \Phi(|\vec k- \vec q|,\tau)}{\cal H}\right)~\nonumber.
\end{eqnarray}
Here $c$ and $\Psi,\Phi$ are defined in Eq.~\eqref{eq:split_modes} and initial conditions for $\Psi,\Phi$ are given in section~\ref{sec:Iso_before_nu_dec}. We have used the relation in Eq.~\eqref{eq:v_psi_phi} to get the last line of the above equation.

{For the case of DRDI, the source term of Eq.~\eqref{eq:eom_h_x} becomes }
\begin{align}
    S_{ij} =& 2\,{\cal P}^{ab}{}_{ij}\Big[2\Phi(\partial_a\partial_b\Psi-\partial_a\partial_b\Phi)-\partial_a\Phi\partial_b\Phi+\partial_a\Psi\partial_b\Psi ~~~~~~~~~(\textrm{DRDI})\notag\\
    &\left.+\partial_a\Phi\partial_b\Psi+\partial_b\Phi\partial_a\Psi+8\pi G a^2 \left(\frac{4}{3}\bar \rho_{\gamma_*}\partial_a v_{\gamma_*}\partial_b v_{\gamma_*}+\bar\rho_{\rm dr} \Pi^{(2)}_{\textrm{dr}\,ab}\right)\right]~,
\end{align}
{We note that this expression can be reduced to Eq.~\eqref{eq:source_woDR} by setting $\Psi=\Phi$ and removing the DR contribution. Since we focus on $R_{\rm dr}\ll 1$ in this study, we only keep the leading terms for $R_{\rm dr}\ll 1$.  According to appendix~\ref{app:Boltzmann_DR}, the leading contribution of ${\Pi}^{(2)}_{\textrm{dr}\,ij}$ in the Fourier space is given as}
\begin{align}
    {\Pi}^{(2)}_{\textrm{dr}\,ij} (\vec{k},\tau)&=-\int \frac{\d^3 \vec{q}}{(2\pi)^3}{q}_i{q}_j \int_0^{\tau} \d \tilde\tau \frac{32}{15} v_{\rm dr}(\vec{q},\tilde\tau) \Phi(\vec k-\vec q,\tilde\tau)
    ~~~~~~~~~~(\textrm{DRDI with}~R_{\rm dr}\ll 1)\notag\\ 
    &\times\left[j_0(k(\tilde\tau-\tau))+\frac{10}{7}j_2(k(\tilde\tau-\tau))+\frac{3}{7}j_4(k(\tilde\tau-\tau))\right]~,
\end{align}
{where  $j_n$ is the spherical Bessel function. Here we only keep the terms that are not vanishing after applying the polarization tensor $e^{ij}_{\vec{k},\lambda}$.}

{For the case of DRDI and $R_{\rm dr}\ll 1$, the source term in Eq.~\eqref{eq:general_source} is now given as}
\begin{align}
    &f(\tau,q,|\vec k- \vec q|) = \frac{1}{2}\Phi(q,\tau)\Phi(|\vec k- \vec q|,\tau)+\frac{1}{2}\Psi(q,\tau)\Psi(|\vec k- \vec q|,\tau)
    \notag\\
    &~~+\frac{1}{2}\left(\Phi(q,\tau)+\frac{\partial_\tau \Psi(q,\tau)}{\cal H}\right)\left(\Phi(|\vec k- \vec q|,\tau)+\frac{\partial_\tau \Psi(|\vec k- \vec q|,\tau)}{\cal H}\right)
    \notag\\
    &~~+{2R_{\rm dr}^2 {\cal H}^2} v_{\rm dr}(q,\tau) v_{\rm dr}(|\vec k-\vec q|,\tau)~~~~~~~~~~~~~~~~~~~~~~~~~~~~~~~~~~~~~~~~~~~~~~~~~~~~~~~~~(\textrm{DRDI})
    \notag\\
    &~~+{R_{\rm dr}}{\cal H}\l(\Phi(q,\tau)+\frac{\partial_\tau\Psi(q,\tau)}{\cal H}\r) v_{\rm dr}(|\vec k-\vec q|,\tau) \notag\\
    &~~+{R_{\rm dr}}{\cal H}v_{\rm dr}(q,\tau)\l(\Phi(|\vec k-\vec q|,\tau)+\frac{\partial_\tau\Psi(|\vec k-\vec q|,\tau)}{\cal H}\r) 
    \notag\\
    &~~-\frac{16}{5}R_{\rm dr}{\cal H}^2 \int_0^\tau d \tilde\tau \,v_{\rm dr}(q,\tilde\tau) \Phi(|\vec k- \vec q|,\tilde\tau)\left[j_0(k(\tilde\tau-\tau))+\frac{10}{7}j_2(k(\tilde\tau-\tau))+\frac{3}{7}j_4(k(\tilde\tau-\tau))\right],\label{eq:f_function}
\end{align}
{ where the second to fifth lines come from the term $\propto \partial_iv_{\gamma*}\partial_j v_{\gamma*}$, while the last term is from $\Pi_{\textrm{dr},ij}^{(2)}$. Since the transfer functions $\Psi,\Phi$ is $\mathcal O(R_{\rm dr})$ and $v_{\rm dr}$ is  $\mathcal O(R_{\rm dr}^0)$ in DRDI, all terms in the above equation are of the same order $\mathcal O(R^2_{\rm dr})$.}

The energy density of gravitational waves relevant for observations is given by~\cite{Domenech:SIGWRe}
\begin{eqnarray}
    \rho_{\rm GW}=\frac{M_{\rm pl}^2}{4a^2}\langle h^{\prime ij}h'_{ij}\rangle,
\end{eqnarray}
where $M_{\rm pl}$ is the Planck mass. This definition is valid when all relevant tensor modes are well inside the horizon, and we have used the approximation $h'\approx k h$. The two-point function in the Fourier space can be written as
\begin{eqnarray}
    \langle h_{\vec k,\lambda} h_{\vec{k}',\lambda}\rangle\equiv \frac{2\pi^2}{k^3}\mathcal P_{h,\lambda}(k)(2\pi)^3\delta^3(\vec k +\vec{k}'),
\end{eqnarray}
where $\mathcal P_{h}(k)$ is the dimensionless power spectrum of $h$. It is common to use the spectral density fraction ($\Omega_{\rm GW}$) defined as 
\begin{eqnarray}\label{eq:GW_k}
    \Omega_{\rm GW}(k)\equiv\frac{1}{\bar\rho}\frac{d \rho_{\textrm{GW}}}{d\ln k}=\frac{k^2}{12\mathcal H^2}\overline  {\mathcal P}_{h}(k),
\end{eqnarray} 
where $\overline{\cal P}_h\equiv\sum_\lambda \overline{\mathcal P}_{h,\lambda} $ and the overline denotes the average over oscillations (see below).

\begin{figure}[!tb]
\centering
\includegraphics[width=0.8\textwidth]{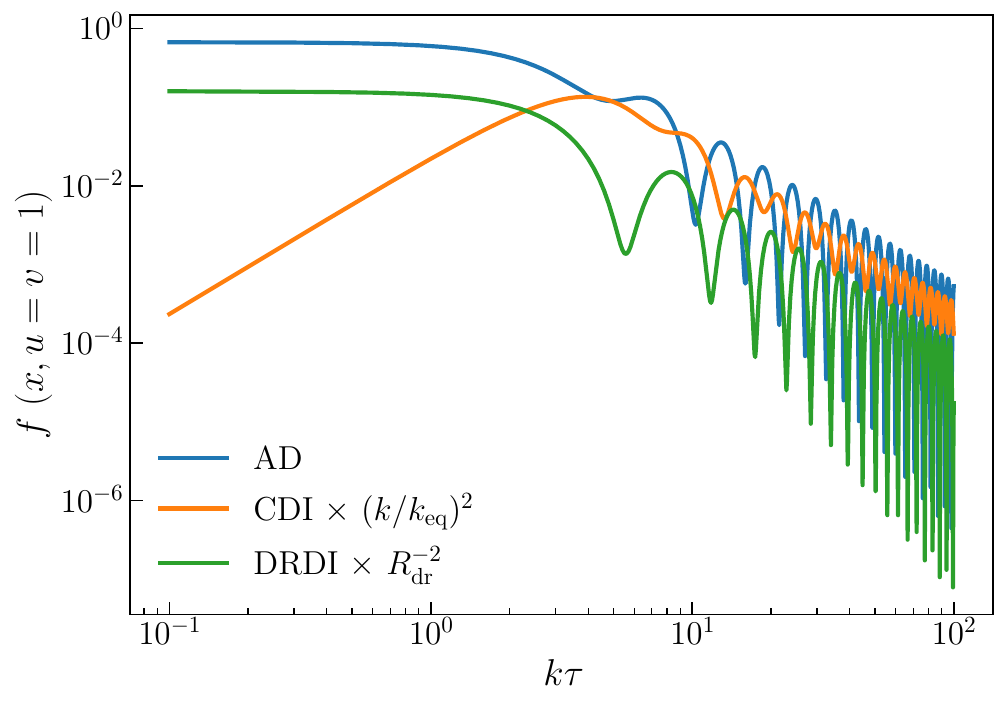}
\caption{The $f(x,u,v)$ function in the source term for induced GWs for AD (blue), CDI (orange) and DRDI (green)  with $x = k\tau$. For simplicity we chose $u = v = 1$ but the main behavior is independent of such particular values.}
\label{fig:f_function}
\end{figure}

The key to obtain GW density is to calculate $\overline{\mathcal P}_h$. Using the Green's function method and assuming no primordial sources of GWs,  we can write the solution of $h_{\vec k,\lambda}$ as
\begin{eqnarray}
    h_{\vec k,\lambda}(\tau) = \int_{0}^{\tau} d \tilde{\tau}\ G(\tau,\tilde{\tau})S_{\vec k,\lambda}(\tilde{\tau})~,
\end{eqnarray}
where $G(\tau,\tilde{\tau}) = k\tilde{\tau}^2[y_0(k\tau)j_0(k\tilde{\tau})-j_0(k\tau)y_0(k\tilde{\tau})]=\sin(k(\tau-\tilde\tau))\times\tilde\tau/(k\tau)$ is the Green's function of Eq.~\eqref{eq:diff_h} in the radiation dominated universe with $j_0$ and $y_0$ being the spherical Bessel functions. We can then write the expression for $\overline{\mathcal P}_h$ as 
\begin{eqnarray}
    \overline{\cal P}_h = 8\int_0^{\infty}d v\int_{|1-v|}^{1+v}d u \left(\frac{4v^2-(1-u^2+v^2)^2}{4uv}\right)^2\overline{I^2(x,v,u)} {\cal P}(ku){\cal P}(kv)~,
    \label{eq:sigw}
\end{eqnarray}
 where we have changed variables $x=k\tau,q=vk, |\vec{k}-\vec{q}|=uk$. $\mathcal P(k)$ is the primordial power spectrum defined in Eq.~\eqref{eq:primordial_P}.
 To get Eq.~\eqref{eq:sigw}, we assume the primordial perturbations are mostly Gaussian, and thus the four-point function of $c$ can be rewritten as the product of two-point functions.

Furthermore, we have inserted all the time dependence into a kernel function defined by
\begin{eqnarray}
    I(x,u,v)= k\int_{0}^{x}d \tilde{x}\ G(x,\tilde{x})f(\tilde{x},u,v)~,
\end{eqnarray}
where $f(x,u,v)$ is given in Eqs.~\eqref{eq:f_function_woDR} or~\eqref{eq:f_function}. Plugging in the Green's function of gravitational waves, the kernel function can be further split as
\begin{eqnarray}
    I(x\gg 1,u,v) = I_y(u,v) y_0(x)+I_j(u,v)j_0(x)~,
\end{eqnarray}
where   
\begin{eqnarray}
    I_{y/j} = \int_0^\infty d \tilde{x} \ \tilde{x}^2 \left\{\begin{matrix} 
  j_0(\tilde{x}) \\  
  y_0(\tilde{x})
\end{matrix}\right\}f(\tilde{x},u,v)~.
\label{eq:Iy/j}
\end{eqnarray}
Since we are interested in the results deep inside the horizon $x= k\tau\gg 1$, $I(x,u,v)$ is a highly oscillatory function due to $y_0(x)$ and $j_0(x)$. To get a meaningful result, we take a time average over oscillations around $x$ and obtain
$\overline {y^2_0(x)}=\overline {j^2_0(x)}=1/(2x^2)$, $\overline {y_0(x)j_0(x)}=0$. This also reflects the fact that the measurement of stochastic GW background can not resolve the real time evolution but just the time-averaged power. Therefore, we can get
\begin{eqnarray}\label{eq:averaged_I}
    \overline{I^2(x,u,v)} =\frac{I_y^2(u,v)+I_j^2(u,v)}{2x^2}.
\end{eqnarray}
 Combining Eqs.~\eqref{eq:GW_k},~\eqref{eq:sigw} and \eqref{eq:averaged_I} as well as $\mathcal H\approx\tau^{-1}$ in radiation domination, we can infer that $\Omega_{\rm GW}(k)$ is independent of $\tau$ around the generation. Taking into account the subsequent heating of the radiation bath from changing degrees of freedom, the observed GW density today is 
\begin{align}
     h^2\Omega_{\rm GW,0}(k)= h^2\Omega_{r,0}\left(\frac{g_*(T_{\rm GW})}{g_{*}(T_0)}\right)\left(\frac{g_{*S}(T_0)}{g_{*S}(T_{\rm GW})}\right)^{4/3}\Omega_{\rm GW}(k)= 3.0\times 10^{-5}\left(\frac{17.25}{g_*(T_{\rm GW})}\right)^{1/3}\Omega_{\rm GW}(k),\nonumber\\
\end{align}
where $h\equiv H_0/(100\,\textrm{km/s/Mpc})$, $g_*(T)$ and $g_{*S}(T)$ are the effective number of relativistic degrees of freedom for energy and entropy, respectively. $T_0$ is the present-day temperature and $T_{\rm GW}$ is the temperature when the GWs are generated, which is $T_{\rm GW}=O(100)$MeV in this study. Therefore,  $g_{*}(T_{\rm GW})=g_{*S}(T_{\rm GW})=17.25$, which includes the contribution from pions, muons, and electrons/positions. The radiation density today $\Omega_{r,0}h^2=4.2\times 10^{-5}$ is taken from~\cite{Planck:2018vyg}.

To explicitly view the effects on induced GWs from different initial conditions, we should look at the source term, in particular $f(x,u,v)$ defined in Eqs.~\eqref{eq:f_function_woDR} or~\eqref{eq:f_function} which encodes all the time evolution. As can be seen from figure~\ref{fig:f_function}, the subhorizon dynamics are rather universal, which exhibit damped oscillations. However, the superhorizon behavior depends on the certain initial condition. Both the adiabatic and DRDI cases remain constant at superhorizon $x\ll1$, whereas the CDI case exhibits a $\tau^2$ dependence. These features originate from the fact that $\Phi$ and $\Psi$ are constant for AD and DRDI but $\propto \tau$ for CDI outside the horizon.   We note that the CDI case is further suppressed by $(k_{\rm eq}/k)^2$ and the DRDI case by $R_{\rm dr}^2$, which reflects the suppression of the density of the corresponding species. Therefore, for modes relevant for PTA observations ($k\gg k_{\rm eq}$), the source term of CDI is highly suppressed but that of DRDI can still be sizable for appropriate $R_{\rm dr}^2$. The cases of BDI and NDI are similar to CDI with an overall rescaling, and thus we don't show them here.

\begin{figure}
\centering
\includegraphics[width=0.45\textwidth]{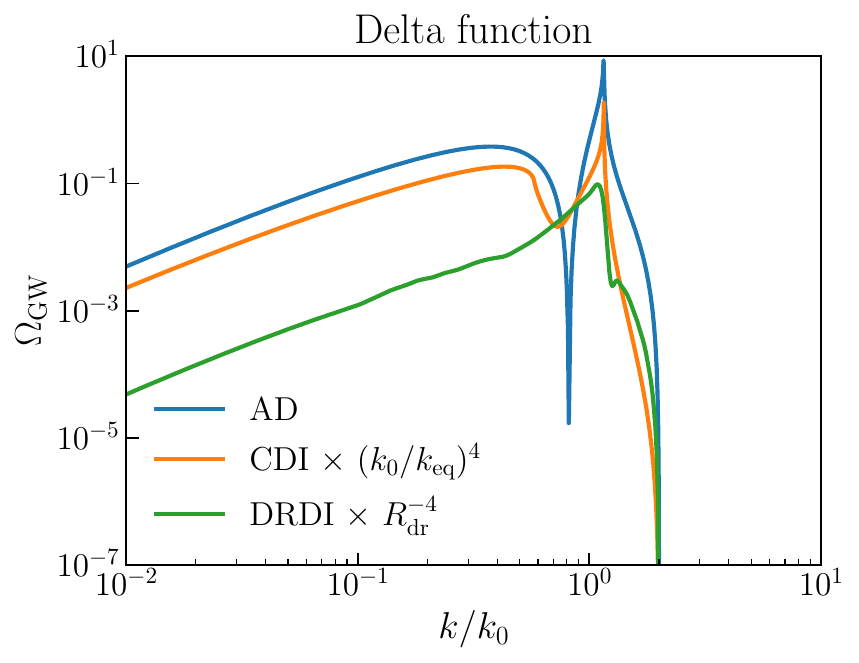}
\includegraphics[width=0.45\textwidth]{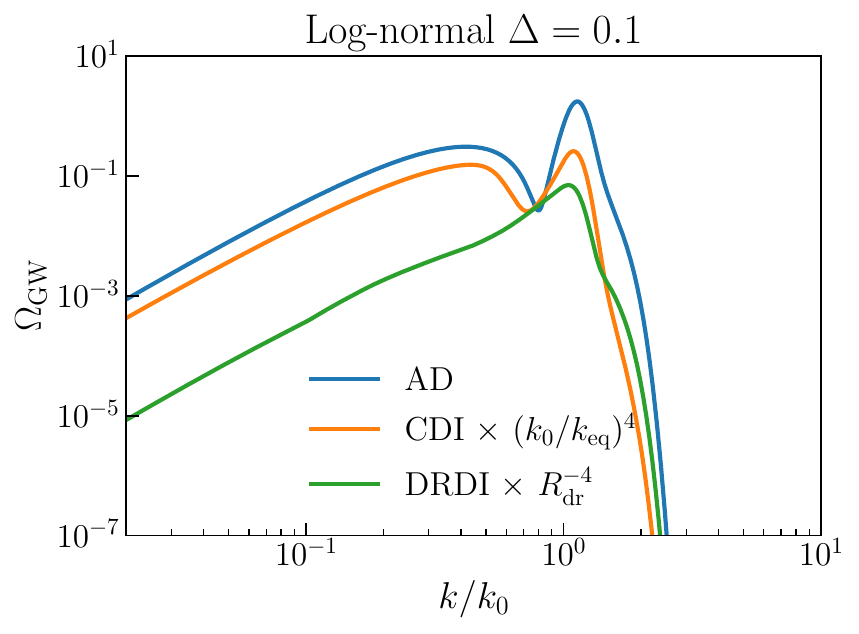}
\includegraphics[width=0.45\textwidth]{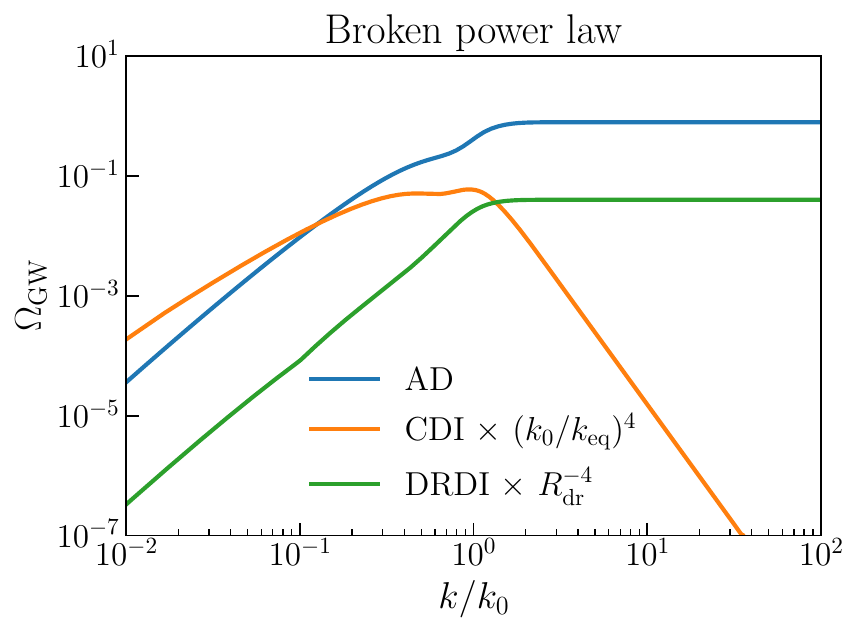}
\caption{The spectrum of GWs ($\Omega_{\rm GW}$) induced by AD (blue), CDI (orange) and DRDI (green) perturbations. We consider three choices of power spectrum: a delta function (Eq.~\eqref{eq:P_delta}), a log-normal form with $\Delta=0.1$ (Eq.~\eqref{eq:P_lognormal}) and a broken power law (Eq.~\eqref{eq:P_Broken_Powerlaw}). For each power spectrum, $k_0$ denotes the characteristic scale and we have set the amplitude $A_{\rm ad/iso}=1$. Here we choose $k_0\gg k_{\rm eq}$ for CDI and $R_{\rm dr}\ll1$ for DRDI. }
\label{fig:GW}
\end{figure}

Given the isocurvature power spectrum $\mathcal P(k)$ in Eqs.~\eqref{eq:P_delta}, \eqref{eq:P_Broken_Powerlaw} and \eqref{eq:P_lognormal}, we can obtain the spectrum of induced GWs ($\Omega_{\rm GW}$). As seen from figure~\ref{fig:GW}, the results for AD and DRDI are almost the same for a given power spectrum, up to an overall coefficient $\sim R_{\rm dr}^4$ for DRDI. This result is consistent with the behavior of $f(x,u,v)$ discussed before. However, there are subtle differences for the delta-function power spectrum. There is a  dip to the left side of the peak of the GW spectrum for AD case, but not for DRDI. This feature has been smeared out for the smooth $\mathcal P(k)$. The CDI case has a very different shape for large $k$, which is the result of suppression of matter density at early times. For the broken power law $\mathcal P(k)$ in the flat region ($k>k_0$), both AD and DRDI cases stay constant but CDI $\propto k^{-4}$. Therefore, the GW spectrum induced by DRDI is dramatically different from the previously studied CDI case. This provides the opportunities to study new GW spectra from isocurvature induced GWs, and the constraints on new physics with the existing GW data.

\paragraph{Curvature vs Isocurvature Induced GWs}

{Here we would like to clarify some intuitions for curvature and isocurvature induced GWs. One common intuition of SIGWs is that GWs are always sourced by metric perturbations (or curvature perturbations). This is indeed the case for previous studies where there is no anisotropic stress ($\Phi=\Psi$) and the source term of GWs is simply a function of $\Phi$ and its derivatives (see Eq.~\eqref{eq:f_function_woDR}). The size of curvature and isocurvature induced GWs can be estimated by the peak value of $\Phi$, which is around the horizon crossing: $\Omega_{\rm GW}\propto \Phi^4|_{k\tau\sim 1}$. Therefore, it is intuitive to understand that GWs induced by isocurvature cases without anisotropic stress (CDI/BDI/NDI) are suppressed (compared to the AD case) because $\Phi^4|_{k\tau\sim 1}\propto (\omega/k)^4$ for $k\gg\omega$. This suppression can also be understood from the fact that the matter density is suppressed at early times.}

{However, this intuition does not capture all interesting features of DRDI induced GWs. Due to the net anisotropic stress from free-streaming DR, the source term cannot be simply rewritten in terms of metric perturbations $\Phi$ and $\Psi$. There are additional terms involving $R_{\rm dr}v_{\rm dr}$~(see Eq.~\eqref{eq:f_function}). These terms contribute the same order as metric perturbations because $\Phi,\Psi $ is $\mathcal O(R_{\rm dr})$ in DRDI with $R_{\rm dr}\ll 1$. Therefore,  the spectrum of DRDI induced GWs  is qualitatively different from other isocurvature cases, and the density of induced GWs can be much larger than those from CDI since DR density ratio ($R_{\rm dr}$) stay fixed in the radiation domination. Although the usual intuition $\Omega_{\rm GW}\propto \Phi^4|_{k\tau\sim1}\propto R_{\rm dr}^4$ (for $R_{\rm dr}\ll 1$) still parametrically holds, it totally obscures the novel contribution from DR density perturbations.  Therefore, one needs to study the full source term to understand induced GWs for general isocurvature cases.}

\section{Constraints from Pulsar Timing Arrays}\label{sec:PTA}
\begin{figure}[!tb]
\centering
\includegraphics[width=0.45\textwidth]{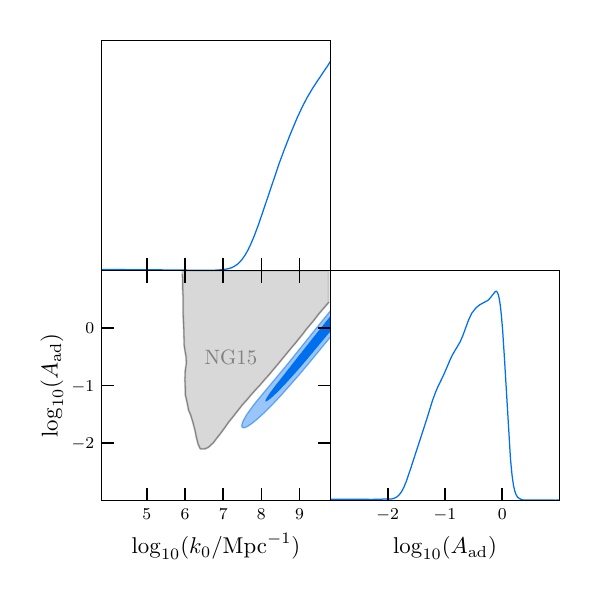}
\includegraphics[width=0.45\textwidth]{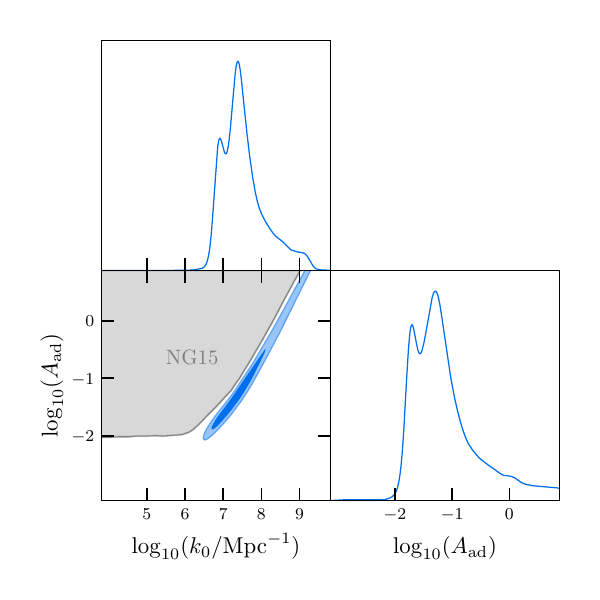}
\caption{The best fit and constraint of AD power spectrum from NANOGrav. We utilize two general parametrizations of the isocurvature power spectrum: a delta function (left, Eq.~\eqref{eq:P_delta}) and a broken power law (right, Eq.~\eqref{eq:P_Broken_Powerlaw}). Each spectrum has a characteristic scale $k_0$ and amplitude $A_{\rm ad}$. The constraint is shown in the gray shaded region, while $1\sigma,\ 2\sigma$ CL regions are depicted in progressively lighter shades of blue.}
\label{fig:constraint_AD}
\end{figure}
Having studied the GW spectrum induced by primordial adiabatic and various isocurvature perturbations, we will perform data fitting with the NANOGrav 15-year dataset~\cite{NANOGrav:2023gor,NANOGrav:2023hde,NANOGrav:2023hvm,nano_grav_2023_dataset}. In this section, we will present best fit and constraints for AD, CDI and DRDI assuming only one mode dominates for each case. Since BDI and NDI have similar observational signals as CDI, their results can be simply recast from CDI with an overall rescaling.  For our data analysis, we employ the parameter $k_0$ and the power spectrum amplitude $A_{\rm ad/iso}$ for each mode. We apply the Bayesian inference method to determine the best fit of SIGWs. We adopt \texttt{PTArcade}~\cite{andrea_mitridate_2023,Mitridate:2023oar} to sample the posterior probability and generate upper limits above which the additional model is “strongly disfavored” according to the Jeffreys scale~\cite{NANOGrav:2023hvm,Jeffreys}. The priors of parameters follow log-uniform distributions within \(\log_{10}(k_0/{\rm Mpc}^{-1})\in [4,10]\),  \(\log_{10}(A_{\rm ad})\in [-3,1]\), \(\log_{10}(R_{\rm dr}^2 A_{\rm iso})\in [-3,1]\), and \(\log_{10}((k_{\rm eq}/k_0)^2A_{\rm iso})\in [-3,1]\).

We first show the constraints and best fit for adiabatic (curvature) perturbations (see figure~\ref{fig:constraint_AD}). Constraints on curvature perturbations from the latest PTA data have been discussed in several recent studies~\cite{NANOGrav:2023hvm,Ellis:2023oxs, Cai:2023dls,Figueroa:2023zhu,Yi:2023tdk,Yi:2023mbm,You:2023rmn,Franciolini:2023pbf,Liu:2023ymk,Liu:2023pau,Liu:2023hpw,Wang:2023ost,Kumar:2024hsi,Chen:2024fir}. Our results with a delta function spectrum agree with those existing results, which serves as the validation of our data analysis. We then present new results for adiabatic perturbations with the broken power law spectrum (see Eq.~\eqref{eq:P_Broken_Powerlaw}). Current data indicate that the amplitude of the adiabatic power spectrum $A_{\rm ad}\lesssim 10^{-2}$ around $k_0\sim 10^6\,\textrm{Mpc}^{-1}$ for the delta function power spectrum, or $k_0\lesssim 10^6\,\textrm{Mpc}^{-1}$
for the broken power law spectrum.

The constraint for CDI and DRDI with the delta function power spectrum exhibit similar features as the AD case. They all have the strongest constraining power around $k_0\sim 10^6\,\textrm{Mpc}^{-1}$, but the sensitivity drops sharply for $k_0\lesssim 10^6\,\textrm{Mpc}^{-1}$ and gradually decreases for $k_0\gtrsim 10^6\,\textrm{Mpc}^{-1}$. This reflects the shape of induced GW spectrum (figure~\ref{fig:GW}), which has a peak around $k_0$, a sharp edge for large $k$ and a long tail for small $k$. The differences are the peak constraint reads $\log_{10}((k_{\rm eq}/k_0)^2A_{\rm iso})\lesssim -2.05$ for CDI and $\log_{10}(R_{\rm dr}^2A_{\rm iso})\lesssim -1.38$ for DRDI.

\begin{figure}[!tb]
\centering
\includegraphics[width=0.45\textwidth]{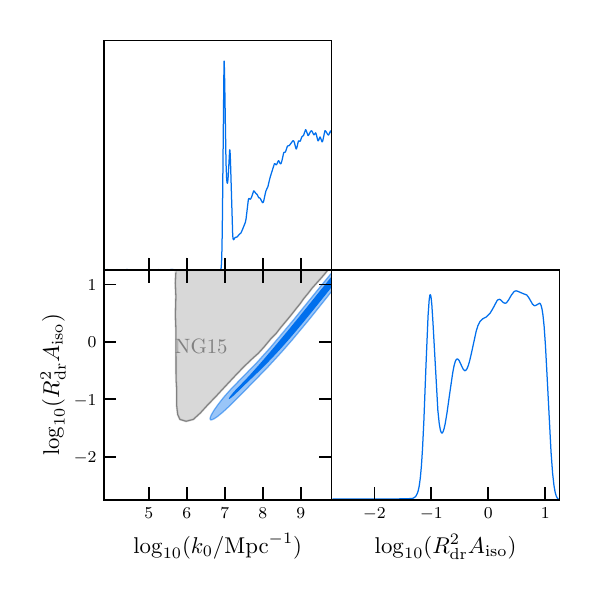}
\includegraphics[width=0.45\textwidth]{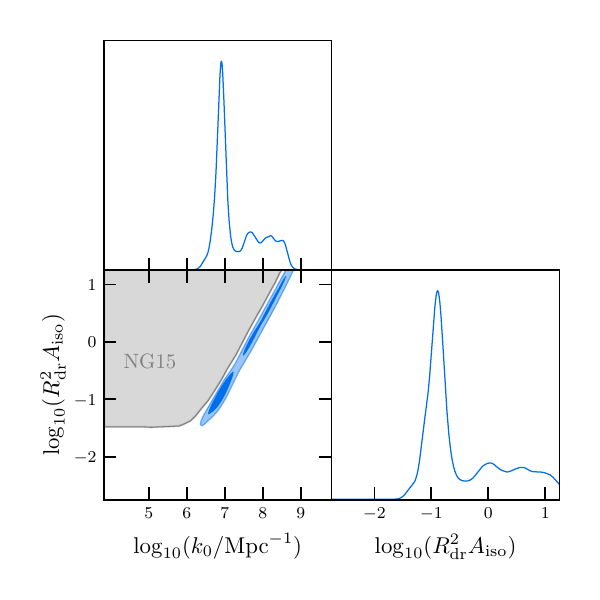}
\includegraphics[width=0.45\textwidth]{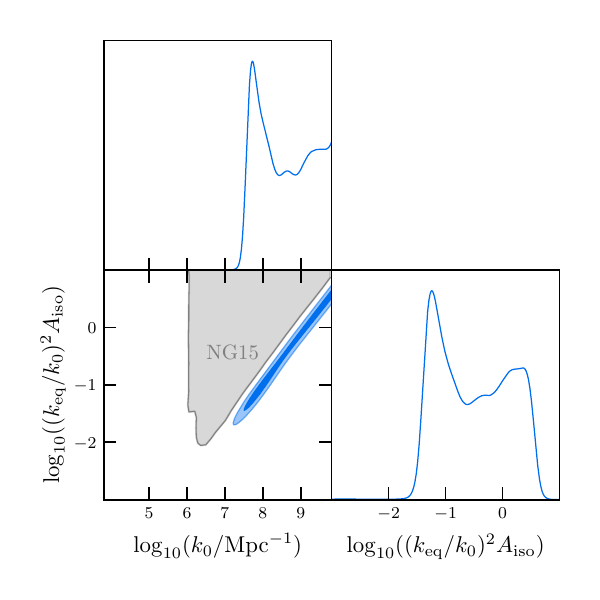}
\includegraphics[width=0.45\textwidth]{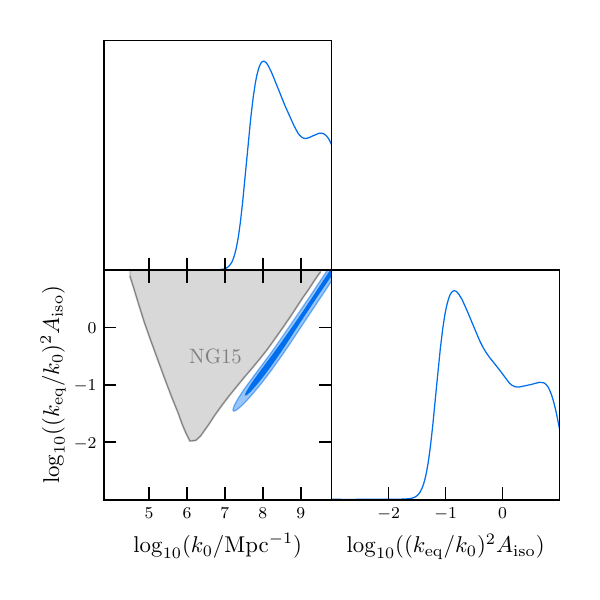}
\caption{The best fit and constraint of DRDI and CDI power spectrum from NANOGrav. We utilize two general parametrizations of the isocurvature power spectrum: a delta function (left, Eq.~\eqref{eq:P_delta}) and a broken power law (right, Eq.~\eqref{eq:P_Broken_Powerlaw}). Each spectrum has a characteristic scale $k_0$ and amplitude $A_{\rm iso}$. The constraint is shown in the gray shaded region, while $1\sigma,\ 2\sigma$ CL regions are depicted in progressively lighter shades of blue.}
\label{fig:constraint_Iso}
\end{figure}

The constraint on DRDI with a broken power law spectrum is flat for $k_0\lesssim 10^6\,\textrm{Mpc}^{-1}$ and weakens for large $k_0$ (see figure~\ref{fig:constraint_Iso}). These features originate from the DRDI GW spectrum (figure~\ref{fig:GW}) which has a plateau for $k\gtrsim k_0$ and decays for $k\lesssim k_0$. However, the CDI constraint on $\log_{10}((k_{\rm eq}/k_0)^2A_{\rm iso})$ has a different low-$k$ behavior that scales as $k_0^{-2}$. This is because the high-$k$ region of the GW spectrum of CDI scales as $k^{-4}$.

\begin{figure}[!tb]
\centering
\includegraphics[width=0.8\textwidth]{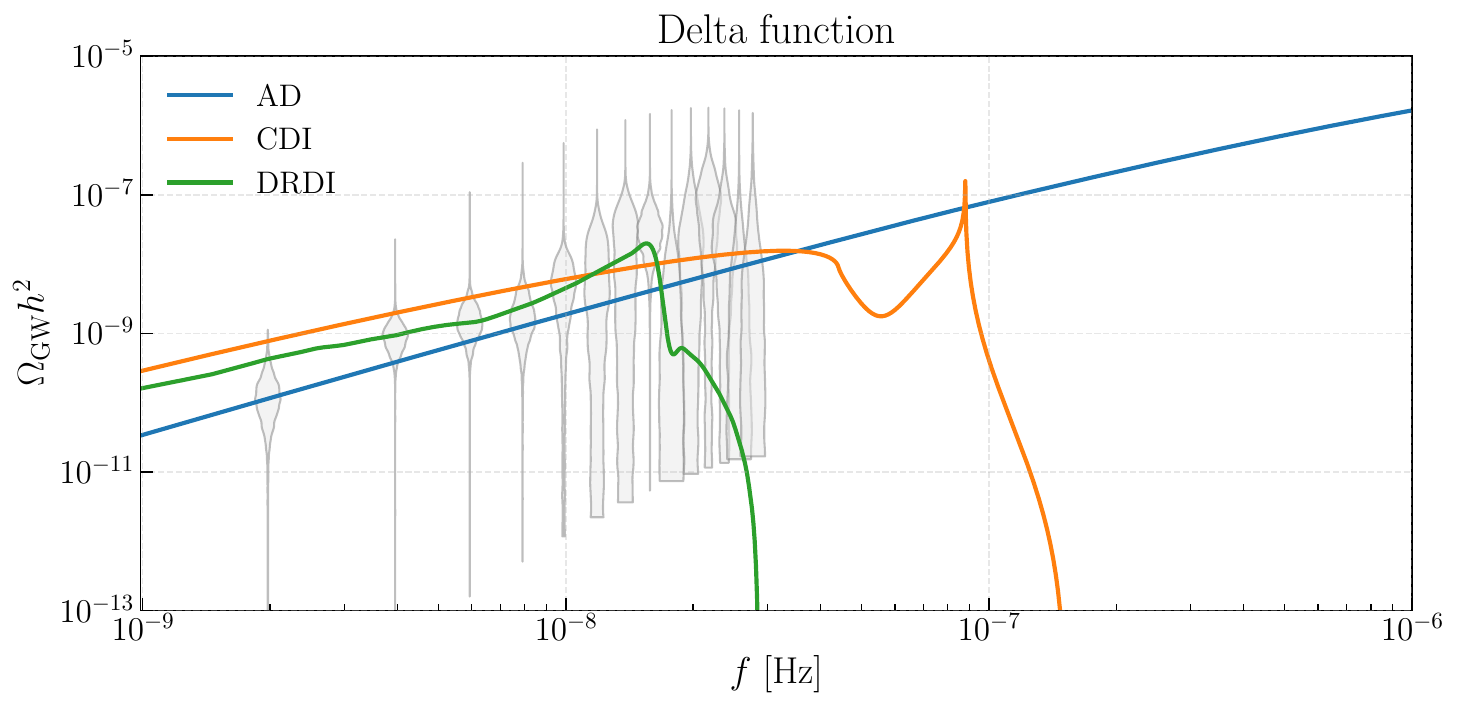}
\includegraphics[width=0.8\textwidth]{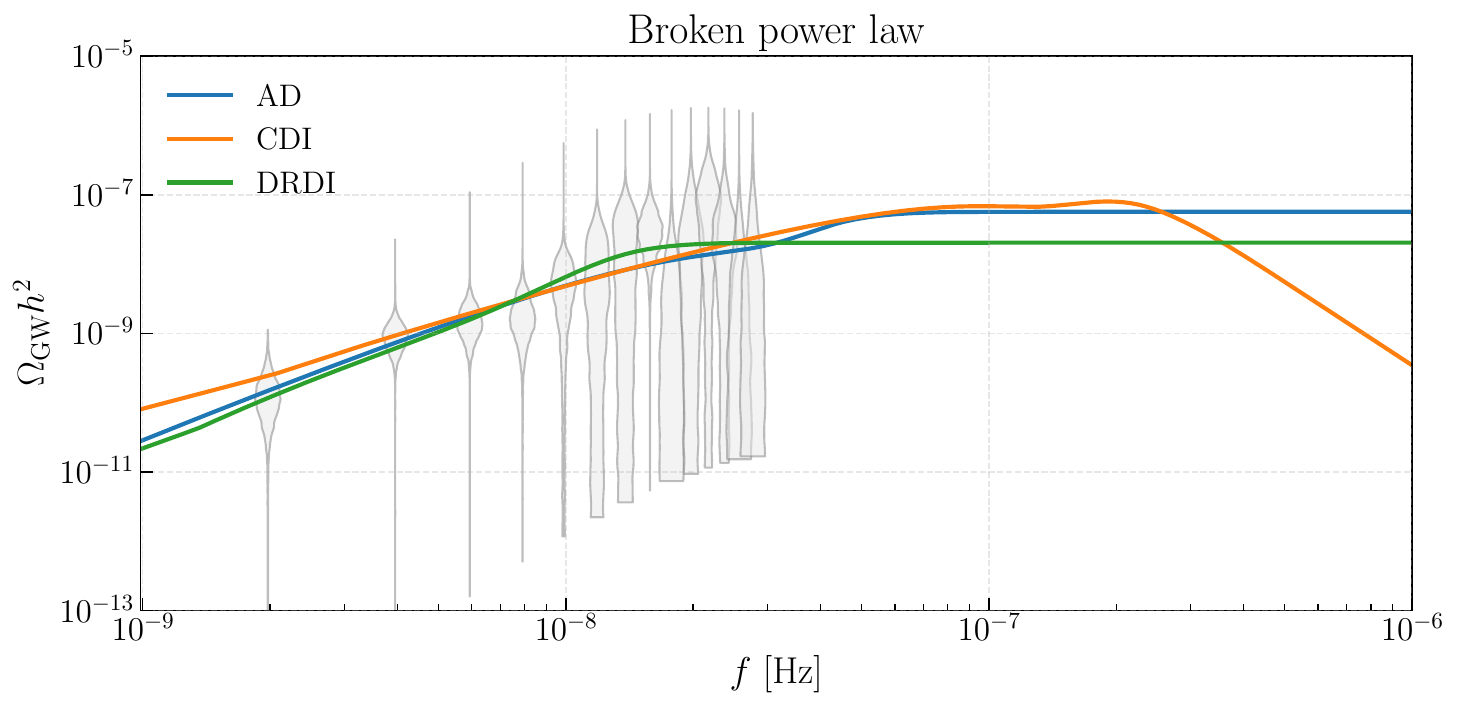}
\caption{The spectrum of induced GWs for the best fit parameters with NANOGrav 15 year data. We show the results for two choices of isocurvature power spectrum: a delta function (top, Eq.~\eqref{eq:P_delta}) and a broken power law (bottom, Eq.~\eqref{eq:P_Broken_Powerlaw}). The data with error bars from NANOGrav are shown in gray. }
\label{fig:nanograv}
\end{figure}

{The $1\sigma$ ($2\sigma$) contours shown in figures~\ref{fig:constraint_AD} and \ref{fig:constraint_Iso} reveal a correlation between the amplitude of the power spectrum and the characteristic scale $k_0$. This behavior reflects the degeneracy between these two parameters when data  are mostly sensitive to the IR portion of the induced GW spectrum. We also show the spectrum of induced GWs (in figure~\ref{fig:nanograv}) for best fit parameters of different cases listed in table~\ref{tab:bestfit}. As seen in this figure, the peaks of the spectrum of AD and CDI cases lie beyond the range of data, meaning the data only probe the IR part of the spectrum. Interestingly, the peak of the DRDI induced GWs sits within the data because the spectrum near the peak captures some interesting features of the data around $f\sim 10^8$ Hz. This difference originates from the peculiar shape of DRDI induced GWs shown in figure~\ref{fig:GW}. For both choices of the isocurvature power spectrum, the data exhibit mild preference for DRDI induced GWs over AD and CDI cases. It will be interesting to study future high precision data to distinguish different isocurvature scenarios. }

\begin{table}[htbp]
  \centering
  
    \begin{tabular}{c|ccc|ccc}
\hline
& \multicolumn{3}{c|}{Delta function} & \multicolumn{3}{c}{Broken power law}  \\\cline{2-7}
          & AD & DRDI & CDI & AD & DRDI & CDI \\
\hline
    $\log_{10}( k_0/{\rm Mpc}^{-1})$ &   9.69    &   6.97   &    7.69   &   7.39   & 6.94  & 8.12 \\
    \textrm{Rescaled Amplitude} &   -0.0523    &   -1.08    &    -1.27   &   -1.30  & -0.879  & -0.668 \\
 \hline
    \end{tabular}%
    \caption{{Best fit parameters of the amplitude of the power spectrum and characteristic scale $k_0$ for AD, CDI and DRDI cases. The value of the rescaled amplitude is $\log_{10}(A_{\rm ad})$ for AD, $\log_{10}(R_{\rm dr}^2A_{\rm iso})$ for DRDI and  $\log_{10}((k_{\rm eq}/k_0)^2A_{\rm iso})$ for CDI.}}
  \label{tab:bestfit}%
\end{table}

\begin{figure}[!tb]
\centering
\includegraphics[width=0.9\textwidth]{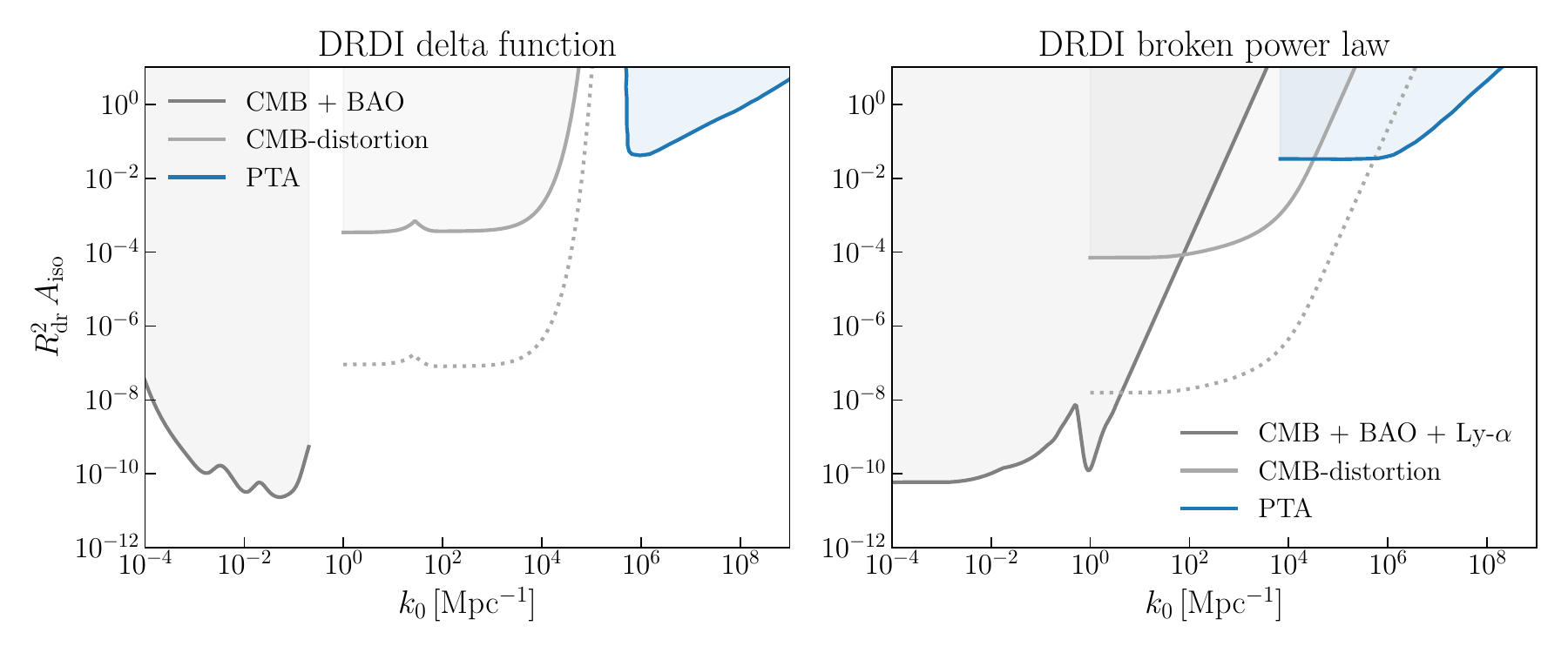}
\includegraphics[width=0.9\textwidth]{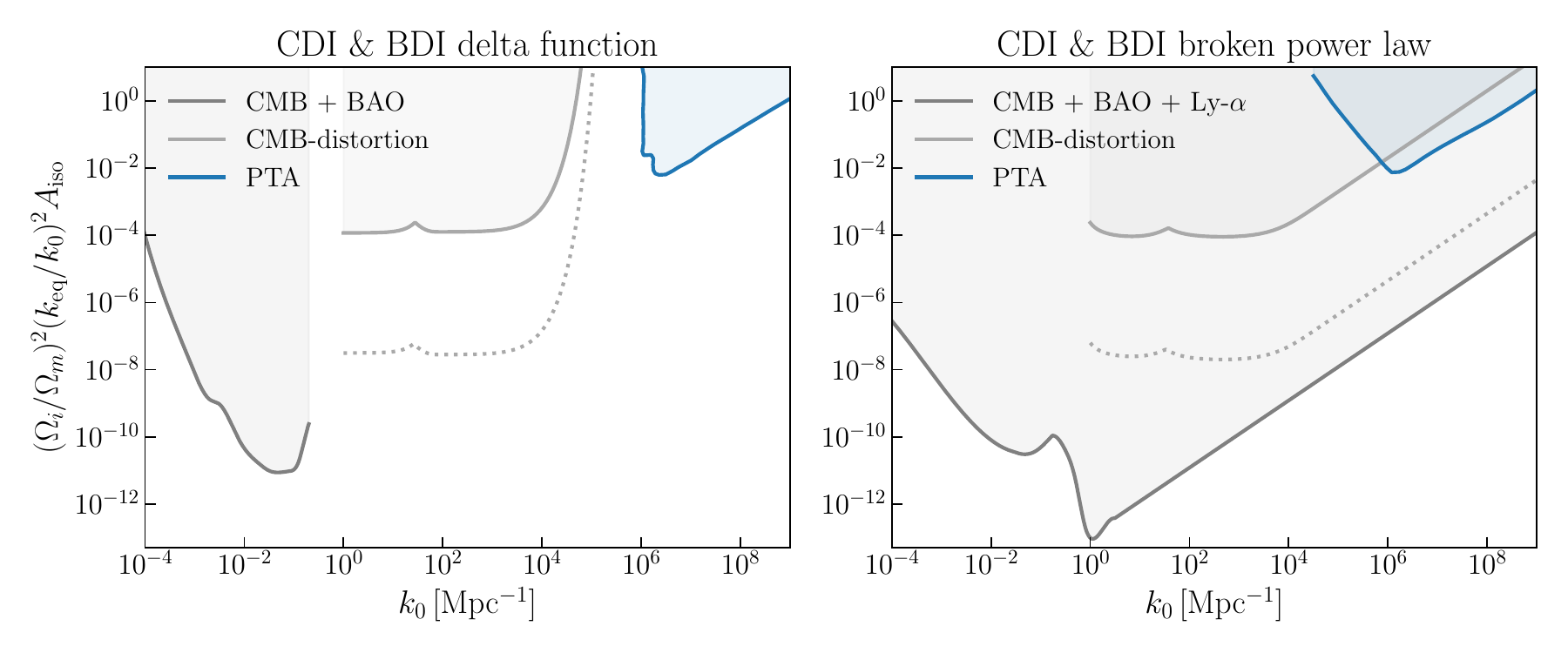}
\caption{Constraints on isocurvature power spectrum for DRDI, CDI and BDI cases with two general parametrizations: a delta function (left, Eq.~\eqref{eq:P_delta}) and a broken power law (right, Eq.~\eqref{eq:P_Broken_Powerlaw}). Our limit from NANOGrav 15 year data is shown in blue. Gray shaded regions are excluded by the CMB, BAO, Lyman-$\alpha$ forest and CMB spectral distortions~\cite{Buckley:2025zgh}. The dotted line indicates the projected sensitivity for the future experiment PIXIE probing CMB distortions~\cite{Kogut:2011xw}. }
\label{fig:compare_limits}
\end{figure}

We further compare the constraints on the isocurvature power spectrum from induced GWs with other observations such as the CMB+BAO, Lyman-$\alpha$ forest and CMB spectral distortions. We can see from figure~\ref{fig:compare_limits} that our constraint from NANOGrav 15 year data places a stringent constraint around $k_0\sim 10^6\textrm{Mpc}^{-1}$, which are complementary to other constraints on larger scales. We acknowledge that a few other strong constraints on isocurvature are not present in this figure because they are derived assuming either BDI or CDI (see e.g.,~\cite{Passaglia:2021jla,Graham:2024hah,Inomata:2018htm,Bagherian:2025puf}).

Finally, we comment on two features of the constraints on isocurvature. Firstly,  the combination $R_{\rm dr}^2 A_{\rm iso}$ for DRDI and $(k_{\rm eq}/k_0)^2A_{\rm iso}$ for CDI controls the physical effect of the perturbation. This means the break-down of the perturbation theory is not $A_{\rm iso}\gtrsim1$, but $R_{\rm dr}^2 A_{\rm iso}\gtrsim 1$ or $(k_{\rm eq}/k_0)^2A_{\rm iso}\gtrsim 1$. For most regions of the constraints derived in this study is safe from this consideration. However, $A_{\rm iso}\gg 1$ means that the isocurvature distribution must be non-Gaussian. The reason is the following. The density contrast $\delta_\alpha\geq -1$ due to $\delta\rho_\alpha+\bar\rho_\alpha\geq0$. If the isocurvature amplitude $A_{\rm iso}\sim \langle\delta_\alpha\delta_\alpha\rangle\gg 1$, the distribution of $\delta_\alpha$ must be mostly around $-1$ to allow for substantial large positive values. Therefore, constraints from our study assuming Gaussian isocurvature fluctuations may need to be modified by non-Gaussian effects given a concrete model.

\section{Conclusions}
\label{sec:conclusions}

The standard $\Lambda$CDM model predicts initial density perturbations are adiabatic, which are consistent with current cosmological observations. However, many well-motivated models of physics beyond the Standard Model can source isocurvature perturbations that leave distinct signatures. Current data can therefore place constraints on the size of isocurvature perturbations. For instance, data from the CMB, BAO, Lyman-$\alpha$ forest, and BBN have set constraints on isocurvature power spectrum across a wide range of scales.
In this study, we focus on another signature of isocurvature from its induced GWs and observations from Pulsar Timing Arrays. We point out that modes relevant for PTA observations enter the horizon before neutrino decoupling, which means standard isocurvature initial conditions derived with free-streaming neutrinos can not apply. We therefore derive new initial conditions with coupled neutrinos, and find an interesting conversion of isocurvature between radiation and matter sectors: NDI results are the same as that of BDI or CDI up to an overall rescaling. Utilizing these new isocurvature initial conditions, we perform a detailed calculation of GWs spectrum induced by different types of isocurvature. The GW spectrum from DRDI is qualitatively different from that of CDI due to the presence of the anisotropic stress. The high-frequency range of the GW spectrum from DRDI is significantly larger than that of CDI due to the density of DR being much larger than DM density at early times.
With two general parametrizations of the isocurvature power spectrum (a delta function and a broken power law), we place constraints using NANOGrav 15 year data. Our results set stringent limits around $10^{6}\,\textrm{Mpc}^{-1}$, which are complementary to other cosmological constraints.

\section*{Acknowledgement}
We thank Lang Liu and Soubhik Kumar for useful discussions on this work and feedback on a draft of this manuscript. This work was supported in part by the National Key R\&D Program of China (2021YFC2203100), by the National Natural Science Foundation of China (12433002, 12261131497), by CAS young interdisciplinary innovation team (JCTD-2022-20), by 111 Project (B23042), by CSC Innovation Talent Funds, by USTC Fellowship for International Cooperation, and by USTC Research Funds of the Double First-Class Initiative.  PD is supported by the National Natural Science Foundation of China (Grants No. T2388102 and 12575111).  JZ is supported by the National Natural Science Foundation of China (125B1023).

\appendix
\section{Evolution Functions for Primordial Perturbations}
In this appendix, we show the set of Einstein equations and Boltzmann
equations that govern the evolution of primordial perturbations. In the Newtonian gauge with the notation defined in section~\ref{sec:Iso_before_nu_dec}, the Einstein equations can be written as~\cite{Kodama:1986fg,Kodama:1986ud,Ma:1995ey}
\begin{eqnarray}
\mathcal{H}^{-1}\Psi^{\prime}+\Phi+\frac{1}{3}\left(\frac{k}{\cal H}\right)^2 \Psi &=& -\frac{1}{2}\delta \\
\frac{k^2}{\mathcal{H}^2}\Psi^{\prime}+\frac{k^2}{\mathcal{H}}\Phi &=& \frac{3}{2}(1+w)\theta \label{eq:v_psi_phi}\\
\mathcal{H}^{-2}\Psi^{\prime\prime}+\mathcal{H}^{-1}\left(\Phi^{\prime}+2\Psi^{\prime}\right)+\left(1+\frac{2{\cal H}'}{{\cal H}^2}\right)\Phi -\frac{1}{3}\left(\frac{k}{\cal H}\right)^2(\Phi-\Psi) &=& \frac{3}{2}\frac{\delta p}{\bar\rho}  \\
\left(\frac{k}{\mathcal{H}}\right)^{2}(\Psi-\Phi) &=& \frac{9}{2}(1+w)\sigma,
\label{eq:Einstein_eq}
\end{eqnarray}
where $w\equiv\bar p/\bar\rho$ and $\delta p\equiv p-\bar p$. We can also relate the total perturbations to the weighted sum of individual ones: $\delta=(\sum_\alpha \bar\rho_\alpha\delta_\alpha)/\bar\rho$, $\theta=(\sum_\alpha (\bar\rho_\alpha+\bar p_\alpha)\theta_\alpha)/(\bar\rho+\bar p)$, $\sigma=(\sum_\alpha (\bar\rho_\alpha+\bar p_\alpha)\sigma_\alpha)/(\bar\rho+\bar p)$. For the cases with no anisotropic stress (AD/BDI/CDI/NDI), we have $\Phi=\Psi$ from the last equation of Eq.~\eqref{eq:Einstein_eq}. Therefore, we only need to solve a single equation to get the evolution for $\Phi$ and $\Psi$.

To obtain the Boltzmann equations for density perturbations, we consider baryons are strongly coupled with the SM radiation bath (denoted as $\gamma_*$) and use the tight-coupling approximation: setting $\theta_b=\theta_{\gamma_*}$ and neglecting the anisotropic stress $\sigma_{\gamma_*}=0$~\cite{Ma:1995ey,Bucher:1999re}. With this approximation, the evolution functions of $\theta_{b}$ and $\theta_{\gamma_*}$ can be combined into one equation. In the radiation dominated era ($w=1/3$), the Boltzmann equations for the density perturbations of different species read
\begin{eqnarray}
\delta_{\gamma_*}^{\prime} &=& -\frac{4}{3} \theta_{\gamma_*}+4\Psi^{\prime}~,\label{eq:fluid_eq} \\
\delta_{c}^{\prime} &=& -\theta_{c}+3\Psi^{\prime}~, \\
\delta_{b}^{\prime} &=& -\theta_{\gamma_*}+3\Psi^{\prime}~, \\
(1+r_b)\theta_{\gamma_*}^{\prime} &=& \frac{1}{4}k^2\delta_{\gamma_*}-r_b{\cal H}\theta_{\gamma_*}+(1+r_b)k^2\Phi~, \\
 \theta_{c}'&=&-\mathcal H \theta_{\rm c}+k^2\Phi~,
\end{eqnarray}
where $r_b\equiv 3\bar\rho_b/(4\bar\rho_{\gamma_*})$. For NDI, we separate $\nu$ from $\gamma_*$ (as mentioned in section~\ref{sec:Iso_before_nu_dec}) but $\nu$ has the same evolution functions as $\gamma_*$, so we don't show it explicitly here.

For DRDI with free-streaming dark radiation, we need to add an additional DR species which contains net anisotropic stress. The complete Boltzmann hierarchy for DR involves an arbitrarily large $\ell$ moment (typically denoted as $F_{\ell}$, see Eq.~\eqref{eq:delta_dr_(n)}):
\begin{eqnarray}
   \delta_{\rm dr}^{\prime} &=& -\frac{4}{3} \theta_{\rm dr}+4\Psi^{\prime}~,\label{eq:Boltzmann_dr} \\
   \theta_{\rm dr}^{\prime} &=& \frac{1}{4}k^2\delta_{\rm dr}-k^2\sigma_{\rm dr}+k^2\Phi~, \\
   F_{\rm dr\,\ell}'&=&\frac{k}{2\ell+1}\left[\ell F_{\rm dr\, \ell-1}-(\ell +1)F_{\rm dr\, \ell+1}\right]~~~~(\ell \geq 2),
\end{eqnarray}
where $\delta=F_0$, $\theta=(3/4)kF_{1}$, $\sigma=(1/2)F_2$.
The effects of high $\ell$ moments on observables are suppressed by powers of $k\tau$ in the superhorizon limit. The Boltzmann hierarchy can therefore be simplified by truncating it at a certain $\ell_{\rm max}$ depending on the desired precision. In this study, we choose 
$\ell_{\rm max}= 8$ and obtain closed differential equations using $F_{\ell_{\rm max}+1}\approx (2\ell_{\rm max}+1) F_{\ell_{\rm max}}/(k\tau)-F_{\ell_{\rm max}-1}$~\cite{Ma:1995ey} and initial conditions $F_{\ell\geq3}=0$.

\section{Isocurvature Initial Conditions After Neutrino Decoupling}
\label{app:Iso_after_nu_dec}
Here we show isocurvature initial conditions after neutrino decoupling, where neutrinos are treated as free-streaming radiation. These cases are widely studied in the literature for the physics relevant for the CMB and large scale structures. The conventional definition of isocurvature is based on setting $\zeta=0$ and a certain $\mathcal S_{\alpha r}\neq0$~\cite{Bucher:1999re}. As argued in section~\ref{sec:Iso_before_nu_dec}, we choose $\mathcal S_{\alpha \gamma}\neq 0$ instead of $\mathcal S_{\alpha r}\neq 0$ to fix isocurvature initial conditions. With this definition, CDI and BDI are the same, but NDI and DRDI are slightly different than those in the literature. We also list the adiabatic initial conditions for comparison. 

In the Newtonian gauge and keeping the leading order in $\tau$, the adiabatic initial conditions are given by
\begin{eqnarray}
       \Psi&=&-\frac{10+4R_\nu}{15+4R_\nu}~~,~~\Phi=-\frac{10}{15+4R_\nu}~ \nonumber\\
    \delta_{c}&=&\delta_{b}=\frac{3}{4}\delta_{\gamma}=\frac{3}{4}\delta_{\nu}=\frac{15}{15+4R_\nu}~~~~~~~~~~~~~~~(\textrm{AD})\nonumber\\
    \theta_{c}&=&\theta_{b}=\theta_{\gamma}=\theta_{\nu}=-\frac{5}{15+4R_\nu}k^2\tau\nonumber\\
    \sigma_\nu&=&\frac{-2}{3(15+4R_\nu)}k^2\tau^2~,
\end{eqnarray}
where $R_\nu\equiv\bar\rho_\nu/\bar\rho_r$. 

The BDI and CDI initial conditions are
\begin{eqnarray}
     \Psi&=&-\frac{1}{8}\frac{15+4R_\nu}{15+2R_\nu}\omega_{b}\tau~,~ \Phi=-\frac{1}{8}\frac{15-4R_\nu}{15+2R_\nu}\omega_{b}\tau~ \nonumber\\
    \delta_c&=&\delta_b-1=\frac{3}{4}\delta_{\gamma}=\frac{3}{4}\delta_{\gamma_\nu}=- \frac{3}{8}\frac{15+4R_\nu}{15+2R_\nu}\omega_{b}\tau ~~~~~~~~~~~~~~~(\textrm{BDI})\nonumber\\
    \theta_{b}&=&\theta_{\gamma}=\theta_{\nu}=-\frac{1}{8}\frac{15}{15+2R_\nu}\omega_{b}k^2\tau^2~,~ \theta_{c}=-\frac{1}{24}\frac{15-4R_\nu}{15+2R_\nu}\omega_{b}k^2\tau^2\nonumber\\
    \sigma_\nu&=& -\frac{1}{6(15+2R_\nu)}\omega_b k^2\tau^3~,
\end{eqnarray}
\begin{eqnarray}
     \Psi&=&-\frac{1}{8}\frac{15+4R_\nu}{15+2R_\nu}\omega_{c}\tau~,~ \Phi=-\frac{1}{8}\frac{15-4R_\nu}{15+2R_\nu}\omega_{c}\tau~ \nonumber\\
    \delta_b&=&\delta_c-1=\frac{3}{4}\delta_{\gamma}=\frac{3}{4}\delta_{\gamma_\nu}=- \frac{3}{8}\frac{15+4R_\nu}{15+2R_\nu}\omega_{c}\tau ~~~~~~~~~~~~~~~(\textrm{CDI})\nonumber\\
    \theta_{b}&=&\theta_{\gamma}=\theta_{\nu}=-\frac{1}{8}\frac{15}{15+2R_\nu}\omega_{c}k^2\tau^2~,~ \theta_{c}=-\frac{1}{24}\frac{15-4R_\nu}{15+2R_\nu}\omega_{c}k^2\tau^2\nonumber\\
    \sigma_\nu&=& -\frac{1}{6(15+2R_\nu)}\omega_c k^2\tau^3~,
\end{eqnarray}
where $\omega_{b/c}\equiv \Omega_{b/c}H_0/\sqrt{\Omega_r}$ and $\Omega_{b/c/r}$ is the fractional energy density of baryons/CDM/radiation today. 

The NDI is given by 
\begin{eqnarray}
     \Psi&=&\frac{R_\nu(1-R_\nu)}{15+4R_\nu}~,~ \Phi=\frac{-2R_\nu(1-R_\nu)}{15+4R_\nu}~ \nonumber\\
     \delta_\gamma&=&\frac{4}{3}\delta_c=\frac{4}{3}\delta_b=-R_\nu\frac{11+8R_\nu}{15+4R_\nu}
     \nonumber\\
     \delta_\nu&=&\frac{(15+8R_\nu)(1-R_\nu)}{15+4R_\nu}~~~~~~~~~~~~~~~~~~~~~~~~~~~~~~~~~~~~~~~~(\textrm{NDI})\nonumber\\
    \theta_{b}&=&\theta_{\gamma}=-\frac{19R_\nu}{4(15+4R_\nu)}k^2\tau~\nonumber\\
    \theta_{\nu}&=&\frac{15(1-R_\nu)}{4(15+4R_\nu)}k^2\tau~,~\theta_{c}=-\frac{R_\nu(1-R_\nu)}{15+4R_\nu}k^2\tau~ \nonumber\\
    \sigma_\nu&=& \frac{(1-R_\nu)}{2(15+4R_\nu)}k^2\tau^2~.
\end{eqnarray}
The DRDI is given by
\begin{eqnarray}
     \Psi&=&-\frac{1}{2}\Phi=\frac{R_{\rm dr}(1-R_{\rm dr}-R_\nu)}{15+4R_{\rm dr}+4R_\nu}~ \nonumber\\
     \delta_\gamma&=&\delta_\nu=\frac{4}{3}\delta_b=\frac{4}{3}\delta_c=-\frac{R_{\rm dr}(11+8R_{\rm dr}+8R_\nu)}{15+4R_{\rm dr}+4R_\nu}\nonumber\\
     \delta_{\rm dr} &=& \frac{15-7R_{\rm dr}-8R_{\rm dr}^2+4R_\nu-8R_{\rm dr}R_\nu}{15+4R_{\rm dr}+4R_\nu}
     \nonumber \\
    \theta_{b}&=&\theta_{\gamma}=\theta_\nu=-\frac{19R_{\rm dr}}{4(15+4R_{\rm dr}+4R_\nu)}k^2\tau~~~~~~~~~~~~~~~~~~~~(\textrm{DRDI})\nonumber\\
    \theta_{c}&=&-\frac{R_{\rm dr}(1-R_{\rm dr}-R_\nu)}{15++4R_{\rm dr}+4R_\nu}k^2\tau~,~\theta_{\rm dr} = \frac{15-15R_{\rm dr}+4R_\nu}{4(15+4R_{\rm dr}+4R_\nu)}k^2\tau \nonumber\\
    \sigma_\nu&=& -\frac{19R_{\rm dr}}{30(15+4R_{\rm dr}+4R_\nu)}k^2\tau^2~,~\sigma_{\rm dr}=\frac{15-15R_{\rm dr}+4R_\nu}{30(15+4R_{\rm dr}+4R_\nu)}k^2\tau^2~.
\end{eqnarray}

We also show these initial conditions in the synchronous gauge. 
 The perturbed FRW metric in the synchronous gauge is written as
\begin{equation}
    ds^2 = a(\tau)^2 \left[-d\tau^2 + (\delta_{ij} + h_{ij})dx^idx^j\right].
\end{equation}
The metric perturbation $h_{ij}$ can be written in Fourier space as
\begin{equation}
   h_{ij}(\vec{k}, \tau) = \left[\hat{k}_i\hat{k}_jh(\vec k, \tau) + \left(\hat{k}_i\hat{k}_j - \frac{1}{3}\delta_{ij}\right)6\eta(\vec k, \tau)\right],
\end{equation}
where $h$ and $\eta$ denote the trace and traceless longitudinal part of $h_{ij}$ respectively. 

In the following, we use \(\tilde X\) to represent perturbations in the synchronous gauge. We set \(\tilde\theta_c\) to zero to fix the synchronous gauge. Then the gauge transformation equations are
\begin{align}
    \tilde\sigma_\alpha &= \sigma_\alpha
    \nonumber\\
    \tilde\theta_\alpha&=\theta_\alpha-\theta_c
    \nonumber\\
\tilde\delta_\alpha+3(1+w_\alpha)\frac{\cal H}{k^2}\tilde \theta&=\delta_\alpha+3(1+w_\alpha)\frac{\cal H}{k^2}\theta
    \nonumber\\
    \tilde h''+6\tilde \eta''+{\cal H}(\tilde h'+6\tilde \eta')&=2k^2 \Phi
    \nonumber\\
    2k^2\tilde\eta-{\cal H}(\tilde h'+6\tilde \eta ')&= 2k^2\Psi~.
\end{align}
\(\tilde{h}\) is determined by the condition \(\tilde\theta_c = 0\) as 
\begin{align}
        \tilde h'&=-2 \tilde \delta_c'~.
\end{align}

Here we keep terms up to $\tau^2$. Terms that are vanishing up to this order, we keep the the leading non-vanishing term.
The AD initial conditions are
\begin{eqnarray}
\tilde h &=& -\frac{1}{2}k^2\tau^2
\nonumber\\
\tilde \eta &=& -1+\frac{5+4R_\nu}{12(15+4R_\nu)}k^2\tau^2
\nonumber\\
\tilde \delta_\gamma&=& \tilde\delta_\nu= \frac{1}{3}k^2\tau^2
\nonumber \\
\tilde\delta_{c}&=&\tilde \delta_b=\frac{1}{4}k^2\tau^2~~~~~~~~~~~~~~~~~~~~~~~~~~~~~~~~(\textrm{AD})
\nonumber \\
\tilde\theta_b &=&\tilde\theta_\gamma= \frac{1}{36}k^4\tau^3~
\nonumber\\
\tilde\theta_\nu &=& \frac{1}{36}\frac{23+4R_\nu}{15+4R_\nu}k^4\tau^3
\nonumber\\
\tilde\sigma_\nu&=&\frac{-2}{3(15+4R_\nu)}k^2\tau^2~.
\end{eqnarray}

The BDI and CDI initial conditions are
\begin{eqnarray}
\tilde h &=& \omega_b \tau-\frac{3}{8}\omega_b\omega\tau^2
\nonumber\\
\tilde\eta &=& -\frac{1}{6}\omega_b\tau+\frac{1}{16}\omega_b\omega\tau^2
\nonumber\\
\tilde\delta_\gamma&=& \tilde\delta_\nu= -\frac{2}{3}\omega_b\tau+\frac{1}{4}\omega_b\omega\tau^2
\nonumber \\
\tilde\delta_{c} &=& \tilde\delta_b-1=-\frac{1}{2}\omega_b\tau+\frac{3}{16}\omega_b\omega\tau^2~~~~~~~~~~~~~~~~~~~~~~(\textrm{BDI})
\nonumber \\
\tilde\theta_b &=&\tilde\theta_\gamma=\tilde\theta_\nu = -\frac{1}{12}\omega_bk^2\tau^2
\nonumber\\
\tilde\sigma_\nu&=& -\frac{1}{6(15+2R_\nu)}\omega_b k^2\tau^3~,
\end{eqnarray}
\begin{eqnarray}
\tilde h &=& \omega_c \tau-\frac{3}{8}\omega_c\omega\tau^2
\nonumber\\
\tilde \eta &=& -\frac{1}{6}\omega_c\tau+\frac{1}{16}\omega_c\omega\tau^2
\nonumber\\
\tilde \delta_\gamma&=& \tilde\delta_\nu= -\frac{2}{3}\omega_c\tau+\frac{1}{4}\omega_c\omega\tau^2
\nonumber \\
\tilde\delta_{b} &=& \tilde\delta_c-1=-\frac{1}{2}\omega_c\tau+\frac{3}{16}\omega_c\omega\tau^2~~~~~~~~~~~~~~~~~~~~~~(\textrm{CDI})
\nonumber \\
\tilde\theta_b &=&\tilde\theta_\gamma=\tilde\theta_\nu = -\frac{1}{12}\omega_ck^2\tau^2
\nonumber\\
    \tilde\sigma_\nu&=& -\frac{1}{6(15+2R_\nu)}\omega_c k^2\tau^3~,
\end{eqnarray}

The NDI is given by
\begin{eqnarray}
\tilde h &=& -\frac{3R_{\nu}}{4}\omega \tau +\frac{9R_{\nu}}{32}\omega^2\tau^2
\nonumber\\
\tilde\eta &=& \frac{R_{\nu}}{8}\omega\tau-\left(\frac{R_{\nu}(1-R_\nu)}{6(15+4R_\nu)}k^2-\frac{3R_{\nu}}{64}\omega^2\right)\tau^2
\nonumber\\
\tilde\delta_\gamma&=& \frac{4}{3}\tilde\delta_b=-R_{\nu}+\frac{1}{2}R_{\nu}\omega \tau +\frac{1}{48}R_\nu(8k^2-9\omega^2)\tau^2
\nonumber \\
\tilde\delta_c &=& -\frac{3}{4}R_{\nu}+\frac{3}{8}R_{\nu}\omega \tau-\frac{9}{64}R_{\nu}\omega^2\tau^2
\nonumber\\
\tilde\delta_{\nu} &=& 1-R_{\nu}+\frac{1}{2}R_{\nu}\omega\tau-\frac{1}{48}R_\nu(8(1-R_\nu)k^2+9R_\nu\omega^2)\tau^2~~~~~~~~~~~~~~~~~~(\textrm{NDI})
\nonumber \\
\tilde\theta_b &=&\tilde\theta_\gamma = -\frac{1}{4}R_{\nu}k^2\tau+\frac{R_\nu(1+3R_b-R_\nu)}{16(1-R_\nu)}\omega k^2\tau^2
\nonumber \\
\tilde\theta_{\nu} &=& \frac{1}{4}(1-R_{\nu})k^2\tau+\frac{1}{16}R_\nu\omega k^2\tau^2~
\nonumber \\
\tilde\sigma_\nu&=& \frac{(1-R_\nu)}{2(15+4R_\nu)}k^2\tau^2~.
\end{eqnarray}

The DRDI is given by
\begin{eqnarray}
\tilde h &=& -\frac{3R_{\rm dr}}{4}\omega \tau +\frac{9R_{\rm dr}}{32}\omega^2\tau^2
\nonumber\\
\tilde\eta &=& \frac{R_{\rm dr}}{8}\omega\tau-\left(\frac{R_{\rm dr}(1-R_{\rm dr}-R_\nu)}{6(15+4R_{\rm dr}+4R_\nu)}k^2-\frac{3R_{\rm dr}}{64}\omega^2\right)\tau^2
\nonumber\\
\tilde\delta_\gamma&=& \tilde\delta_\nu=\frac{4}{3}\tilde\delta_b= -R_{\rm dr}+\frac{1}{2}R_{\rm dr}\omega \tau+\frac{1}{48}R_{\rm dr}(8k^2-9\omega^2)\tau^2 
\nonumber \\
\tilde\delta_c &=& -\frac{3}{4}R_{\rm dr}+\frac{3}{8}R_{\rm dr}\omega \tau-\frac{9}{64}R_{\rm dr}\omega^2\tau^2
\nonumber\\
\tilde\delta_{\rm dr} &=& 1-R_{\rm dr}+\frac{1}{2}R_{\rm dr}\omega\tau-\frac{1}{48}R_{\rm dr}(8(1-R_{\rm dr})k^2+9R_{\rm dr}\omega^2)\tau^2~~~~~~~~~~~~~~~~~~(\textrm{DRDI})
\nonumber \\
\tilde\theta_b &=&\tilde\theta_\gamma = \tilde\theta_\nu = -\frac{1}{4}R_{\rm dr}k^2\tau+\frac{R_{\rm dr}(1+3R_b-R_{\rm dr}-R_\nu)}{16(1-R_{\rm dr}-R_\nu)}\omega k^2\tau^2
\nonumber \\
\tilde\theta_{\rm dr} &=& \frac{1}{4}(1-R_{\rm dr})k^2\tau+\frac{1}{16}R_{\rm dr}\omega k^2\tau^2
\nonumber\\
\tilde\sigma_\nu&=& -\frac{19R_{\rm dr}}{30(15+4R_{\rm dr}+4R_\nu)}k^2\tau^2~,~\tilde\sigma_{\rm dr}=\frac{15-15R_{\rm dr}+4R_\nu}{30(15+4R_{\rm dr}+4R_\nu)}k^2\tau^2~.
\end{eqnarray}

\section{Second-order Perturbations of Free-streaming Dark Radiation}\label{app:Boltzmann_DR}

The energy momentum tensor for a certain species can be determined by its phase space distribution $f
(\tau,q,\vec{x},\vec{n})$, where $\tau$ is the conformal time, $\vec{x}$ is the spatial coordinate, $\vec{q}=q\vec{n}$ is the comoving momentum with $q$ being the magnitude and $\vec{n}$ being the unit vector ($\delta_{ij} n^{i}n^{j}=1$). The time evolution of $f$ is controlled by the Boltzmann equation
\begin{eqnarray}
    \frac{d}{d \tau}f=C[f],
\end{eqnarray}
where $d/d\tau$ is the total derivative with respect to $\tau$ and $C[f]$ denotes the collision term. For free-streaming dark radiation (DR), the collision term vanishes. Therefore, the Boltzmann equation for DR is written as
\begin{eqnarray}
    \frac{\partial f_{\rm dr}}{\partial \tau}
    + \frac{\partial f_{\rm dr}}{\partial x^{i}} \cdot \frac{dx^{i}}{d\tau}
    + \frac{\partial f_{\rm dr}}{\partial q} \cdot \frac{dq}{d\tau}
    + \frac{\partial f_{\rm dr}}{\partial n^{i}} \cdot \frac{dn^{i}}{d\tau}
    = 0~.
\end{eqnarray}
The phase space distribution of DR \(f_{\rm dr}\) (up to the second order) can be decomposed as
\begin{align}
    f_{\rm dr}(\tau,q,\vec{x},\vec{n}) = \bar f_{\rm dr}(q)+f^{(1)}_{\rm dr}(\tau,q,\vec{x},\vec{n})+f^{(2)}_{\rm dr}(\tau,q,\vec{x},\vec{n})~.
\end{align}
We consider the unperturbed $\bar f_{\rm dr}$ is the equilibrium distribution (e.g., Fermi-Dirac distribution) which is only a function of $q$. The superscript $(n)$ denotes the perturbation of the $n$-th order. Similarly, metric perturbations in Eq.~\eqref{eq:perturbed_metric} can be written as $\Phi=\Phi^{(1)}+\Phi^{(2)}$,$\Psi=\Psi^{(1)}+\Psi^{(2)}$, $h_{ij}=h_{ij}^{(2)}$. Here we consider the tensor fluctuation is sourced by the product of first-order scalar perturbations, which appears at the second order.

The Boltzmann equation governs the evolution of the first-order perturbation reads
\begin{eqnarray}
    \frac{\partial f_{\rm dr}^{(1)}}{\partial \tau}
        +  \frac{\partial f_{\rm dr}^{(1)}}{\partial x^{i}} n^{i}=-q \frac{\partial \bar f_{\rm dr}}{\partial q}\left(-\frac{\partial\Phi^{(1)}}{\partial x^j}n^j+\Psi^{(1)\prime}\right),
\end{eqnarray}
where the prime denotes the partial derivative with respect to $\tau$. The general solution to $f_{\rm dr}^{(1)}$ in the Fourier space can be obtained as
\begin{eqnarray}\label{eq:fdr_(1)}
    f_{\rm dr}^{(1)}(\tau,q,\vec{k},\vec{n})&=&C(\vec{k},q) e^{-i \vec{k} \cdot\vec{n}\, \tau}\\
    &+&q \frac{\partial \bar f_{\rm dr}}{\partial q}\left(\Phi^{(1)}(\vec{k,\tau})-\int_0^{\tau}d\tilde\tau\, (\Phi^{(1)\prime}(\vec{k,\tau})+\Psi^{(1)\prime}(\vec{k,\tau}))e^{-i \vec{k}\cdot\vec{n}(\tau-\tilde\tau)}\right),\nonumber
\end{eqnarray}
where the first line comes from the solution without metric perturbations and $C(\vec{k},q)\approx \bar f_{\rm dr}(q)c(\vec{k})$ for $R_{\rm dr}\ll 1$ is determined by the definition of density perturbations in Eqs.~\eqref{eq:split_modes} and \eqref{eq:transfer_DRDI}. For DRDI with $R_{\rm dr}\ll 1$, the first line dominates because $\Phi^{(1)},\Psi^{(1)}$ vanishes as $R_{\rm dr}\to 0$. Therefore, we will only keep the first line for later analysis.

The Boltzmann equation for the second-order perturbation $f_{\rm dr}^{(2)}$ is~\cite{Mangilli:2008bw,Saga:2014jca}
\begin{align}
    \frac{\partial f_{\rm dr}^{(2)}}{\partial \tau}
        +  \frac{\partial f_{\rm dr}^{(2)}}{\partial x^{i}} n^{i}
        = & -\frac{\partial f_{\rm dr}^{(1)}}{\partial x^{i}} n^{i} \left( \Phi^{(1)} + \Psi^{(1)} \right)  
        - q \frac{\partial f_{\rm dr}^{(1)}}{\partial q} \left( -n^{j} \frac{\partial\Phi^{(1)}}{\partial x^j} + \Psi^{(1)\prime} \right) \notag\\
        & - \frac{\partial f_{\rm dr}^{(1)}}{\partial n^{i}} \left[ n^{i} n^{j} \frac{\partial}{\partial x^j}\left( \Phi^{(1)} + \Psi^{(1)} \right) - \frac{\partial}{\partial x_i}\left(\Psi^{(1)} + \Phi^{(1)} \right)\right] +...\label{eq:2ndBoltzmann}
\end{align}
where we have neglected terms that vanish after applying the polarization tensor $e^{ij}_{\vec{k},\lambda}$ because they do not source GWs. We have also dropped the term involving $h^{(2)}_{ij}$ because it leads to a damping term which is further suppressed by $R_{\rm dr}$~\cite{Weinberg:2003ur,Mangilli:2008bw,Zhang:2022dgx}.

It is useful to define the brightness function of DR
\begin{align}\label{eq:delta_dr_(n)}
    F^{(n)}_{\rm dr} = \frac{\int q^3\d q f^{(n)}_{\rm dr}}{\int q^3 \d q \bar f_{\rm dr}}~.
\end{align}
We note that the definition of $F_{\rm dr}^{(1)}$ matches the usual definition of $F_{\rm dr}$ in Ref.~\cite{Ma:1995ey}. Since the dependence of $\vec{n}$ in first-order perturbations is only through $\vec{\hat k}\cdot\vec{n}$~\cite{Ma:1995ey}, we can expand it using Legrendre polynomials $P_{\ell}$ as
\begin{eqnarray}
    F^{(1)}_{\rm dr}(\tau,\vec k,\vec{\hat k}\cdot\vec{n}) =\sum_\ell^\infty (-i)^\ell(2\ell+1) F_{\textrm{dr}\,\ell}^{(1)}(\tau,\vec{k}) P_\ell (\hat{\vec k}\cdot \vec{n}).
\end{eqnarray}
Plugging Eq.~\eqref{eq:fdr_(1)} into Eq.~\eqref{eq:delta_dr_(n)}, we obtain
\begin{eqnarray}
    F^{(1)}_{\rm dr} (\tau,\vec{k}, \vec{\hat k}\cdot\vec{n}) = c(\vec{k})e^{-i\vec{k} \cdot\vec{n}\, \tau}~;~F^{(1)}_{\textrm{dr}\,\ell} (\tau,\vec{k})=c(\vec{k})j_\ell (k\tau)~~~~(\textrm{DRDI~with}~R_{\rm dr}\ll 1),
\end{eqnarray}
where $c(\vec{k})$ is defined in Eq.~\eqref{eq:split_modes}.

To get the second-order perturbation $F^{(2)}_{\rm dr}$, we replace \(f^{(2)}_{\rm dr}\) with \(F^{(2)}_{\rm dr}\) in Eq.~\eqref{eq:2ndBoltzmann} and take the Fourier transform: 
\begin{align}
    F_{\rm dr}^{(2)\prime}(\tau,\vec{k},\vec{n})+i\vec{k}\cdot\vec{n} F^{(2)}_{\rm dr}(\tau,\vec{k},\vec{n})=S_{\rm dr}^{(2)}(\tau,\vec k,\vec{n})~,
\end{align}
where 
\begin{align}
    S^{(2)}_{\rm dr}(\tau,\vec{k},\vec{n}) &= \int \frac{\d^3 \vec{q}}{(2\pi)^3}\l[-(i\vec{q}\cdot \vec n)F^{(1)}_{\textrm{dr}\,\vec{q}}(\Phi^{(1)}_{\vec{k}_2}+\Psi^{(1)}_{\vec{k}_2})+4F^{(1)}_{\textrm{dr}\,\vec{q}}\l((-i\vec k_2\cdot \vec n)\Phi^{(1)}_{\vec{k}_2}+\Psi^{(1)\prime}_{\vec{k}_2}\r)\r. \notag
    \\
    &\quad\quad\quad\quad \l.-\frac{\partial F^{(1)}_{\textrm{dr}\,\vec{q}}}{\partial n^j}\l(n^j(i\vec k_2\cdot \vec n)(\Phi^{(1)}_{\vec{k}_2}+\Psi^{(1)}_{\vec{k}_2})-ik_2^j(\Phi^{(1)}_{\vec{k}_2}+\Psi^{(1)}_{\vec{k}_2})\r) \r]~,\label{eq:boltzmannSource}
\end{align}
where \(\vec{k}_2=\vec{k}-\vec{q}\). Since the angular dependence of $F^{(2)}_{\rm dr}$ is controlled by two degrees of freedom $\vec{n}$,  we shall expand it by spherical harmonics $Y_{\ell,m}$ instead of $P_\ell$:
\begin{eqnarray}
    F_{\rm dr}^{(2)}(\tau,\vec{k},\vec{n})=\sum_\ell\sum^{\ell}_{m=-\ell} F_{\textrm{dr}\,\ell,m}^{(2)}(\tau,\vec{k})(-i)^\ell\sqrt{\frac{4\pi}{2\ell+1}}Y_{\ell, m}(\vec{n}).
\end{eqnarray}
As we shall see later, the source term of induced gravitational waves depends on $F_{\textrm{dr}\, 2,\sigma}^{(2)}$ with $\sigma=\pm 2$ denoting two polarizations. According to Eq.~\eqref{eq:boltzmannSource}, the solution of $F_{\textrm{dr}\, 2,\sigma}^{(2)}$ is
\begin{eqnarray}\label{eq:F_dr^2_2}
     F^{(2)}_{\textrm{dr}\,2,\sigma}(\tau,\vec{k}) = \int_0^{\tau} \d \tilde\tau\  S^{(2)}_{\textrm{dr}\,2,\sigma}(\tilde\tau,\vec{k}) \l[j_0(k(\tilde\tau-\tau))+\frac{10}{7}j_2(k(\tilde\tau-\tau))+\frac{3}{7}j_4(k(\tilde\tau-\tau))\r], 
\end{eqnarray}
where 
\begin{eqnarray}
     S_{\textrm{dr}\,2,\sigma}^{(2)}(\tau,\vec{k}) = - \sqrt{\frac{5}{4\pi}}\int d\Omega_n \ S^{(2)}_{\rm dr}(\tau,\vec k,\vec{n}) Y^*_{2,\sigma} (\vec{n}) .
\end{eqnarray}

With useful math identities and angular integrals in section~\ref{sec:useful_math}, we can obtain 
\begin{eqnarray}\label{eq:S_dr_2,2}
      S^{(2)}_{{\rm dr}\,2,\sigma} = \int \frac{d^3 \vec{q}}{(2\pi)^3}   q_{i} q_{j} \sqrt{\frac{2}{3}} e^{ij}_{\vec{k},-\sigma}(16v_{{\rm dr}\,\vec{q}}\Phi_{\vec{k}_2})~~~~(\textrm{DRDI~with}~R_{\rm dr}\ll 1),
\end{eqnarray}
where $-kv_{\rm dr}=(3/4)F^{(1)}_{\textrm{dr}\,1}$ and  $e^{ij}_{\vec{k},\sigma}$ is the polarization tensor that projects the $m=\sigma$ mode.
We used $n_i n_j e^{ij}_{\vec{k},-\sigma} = \sqrt{\frac{2}{3}}\sqrt{\frac{4\pi}{5}} Y_{2,\sigma}^*(\vec n)$, where we choose the $z$ direction parallel to $\vec{k}$ and two polarization vectors as 
\begin{eqnarray}
    \vec e=\frac{1}{\sqrt{2}} (\vec{\hat x}+i\vec{\hat y})~,~\vec{\bar e}=\frac{1}{\sqrt{2}} (\vec{\hat x}-i\vec{\hat y})
\end{eqnarray}
\begin{eqnarray}
    e^{ij}_{\vec k,2}=e^i(\vec{k})e^j(\vec{k})~,~
    e^{ij}_{\vec k,-2}=\bar e^i(\vec k)\bar e^j(\vec k)
\end{eqnarray}
We note that we use the polarization tensor $e^{ij}_{\vec{k},\sigma}$ here for convenience in the calculation with spherical harmonics. However, in the main text we define GWs with $\lambda=+,\times$ modes~(see Eq.~\eqref{eq:general_source})
\begin{eqnarray}
    e^{ij}_{\vec k,+}=\frac{1}{\sqrt 2}\left(e^i(\vec{k})e^j(\vec{k})-\bar e^i(\vec k)\bar e^j(\vec k)\right) \nonumber\\
    e^{ij}_{\vec k,\times}=\frac{1}{\sqrt 2}\left(e^i(\vec{k})\bar e^j(\vec{k})+\bar e^i(\vec k) e^j(\vec k)\right) 
\end{eqnarray}

\subsection{Second-order Energy Momentum Tensor}
The energy-momentum tensor of DR can be written in terms of the phase space distribution function $f_{\rm dr}$ 
\begin{align}
    T^{\;\mu}_{\textrm{dr}\, \nu} = g_{\rm dr} \int \frac{\d P^3}{(2\pi)^3} (-g)^{-1/2} \frac{P_\mu P_\nu}{P^0}f_{\rm dr},
\end{align}
where $g_{\rm dr}$ denotes the degree of freedom of DR.
The second-order perturbation $\delta T^{(2)}_{\rm dr}{}_j^{i}$ is then given as~\cite{Mangilli:2008bw,Zhang:2022dgx}
\begin{align}
    \delta T^{(2)}_{\rm dr}{}_j^{i} = \frac{a^{-4}g_{\rm dr}}{(2\pi)^3}\int d\Omega_n\ q^2 dq\ q\ n^in_jf^{(2)} _{\rm dr}= \frac{\bar\rho_{\rm dr}}{4\pi}  \int d\Omega_n n^i n_j F^{(2)}_{\rm dr},
\end{align}
where $\bar\rho_{\rm dr}\equiv a^{-4}g_{\rm dr}/(2\pi)^3 \int d^3\vec{q}\, q\bar f_{\rm dr}(q)$ and $F^{(2)}_{\rm dr}$ is defined in Eq.~\eqref{eq:delta_dr_(n)}. Using the definition in Eq.~\eqref{eq:T_dr_(2)} of the main text, we obtain
\begin{eqnarray}
\Pi_{\textrm{dr}\,ij}^{(2)}=\int\frac{d\Omega_{n}}{4\pi} n_i n_j F_{\rm dr}^{(2)}.
\end{eqnarray}

The source term for GWs from DR component involves a term that is proportional to $\Pi_{\textrm {dr}\,ij}^{(2)} e^{ij}_{\vec{k},-\sigma}$, which is given as   
\begin{align}\label{eq:Pi_e_(2)}
   \Pi_{\textrm {dr}\,ij}^{(2)} e^{ij}_{\vec{k},-\sigma} =-\frac{1}{5} \sqrt{\frac{2}{3}} F^{(2)}_{{\rm dr}\,2,\sigma}~,
\end{align}
where we have used the relation \(n_i n_j e^{ij}_{\vec{k},\mp 2} = \sqrt{\frac{2}{3}}\sqrt{\frac{4\pi}{5}} Y_{2,\pm 2}^*(\vec n)\) and 
\begin{align}
F^{(2)}_{{\rm dr}\,2,\sigma}(\tau,\vec{k}) = \int d\Omega_n (-1)\sqrt{\frac{5}{4\pi}}Y_{2,\sigma}^*(\vec{n}) F^{(2)}_{\rm dr}(\tau,\vec{k},\vec{n})~.
\end{align}
Comparing Eqs.~\eqref{eq:F_dr^2_2} and \eqref{eq:Pi_e_(2)}, we can obtain
\begin{align}
    {\Pi}^{(2)}_{\textrm{dr}\,ij} (\vec{k},\tau)&=-\int \frac{\d^3 \vec{q}}{(2\pi)^3}{q}_i{q}_j \int_0^{\tau} \d \tilde\tau \frac{32}{15} v_{\rm dr}(\vec{q},\tilde\tau) \Phi(\vec k-\vec q,\tilde\tau)
    ~~~~~~~~~~(\textrm{DRDI with}~R_{\rm dr}\ll 1)\notag\\ 
    &\times\left[j_0(k(\tilde\tau-\tau))+\frac{10}{7}j_2(k(\tilde\tau-\tau))+\frac{3}{7}j_4(k(\tilde\tau-\tau))\right]~.
\end{align}

\subsection{Useful Math Equations}
 \label{sec:useful_math}
In this section, we list a few useful math equations that relevant for calculation in this appendix.

Relations between Legrendre Polynomials with spherical harmonics:
\begin{align}
  P_\ell (\hat{\vec k}\cdot \vec{n})  =  \frac{4\pi}{2\ell+1} \sum_{m=-\ell}^\ell Y_{\ell,m}(\vec{n}) Y_{\ell,m}^*(\hat {\vec{k}}).
\end{align}

Orthonormal conditions for spherical harmonics:
\begin{align}
    \int \d \Omega_n Y_{\ell_1,m_1}^*(\vec{n})Y_{\ell_2,m_2}(\vec{n}) = \delta_{\ell_1\ell_2} \delta_{m_1m_2} ~.
\end{align}

Integrals of the product of three spherical harmonics (Gaunt coefficients):
\begin{align}
    &\int d \Omega_n \, Y_{\ell_{1} m_{1}}({\mathbf{n}}) Y_{\ell_{2} m_{2}}({\mathbf{n}}) Y_{\ell_{3} m_{3}}({\mathbf{n}}) \notag\\
= &\sqrt{\frac{(2\ell_{1}+1)(2\ell_{2}+1)(2\ell_{3}+1)}{4\pi}} \times 
\left(\begin{array}{ccc}
\ell_{1} & \ell_{2} & \ell_{3} \\
0 & 0 & 0
\end{array}\right)
\left(\begin{array}{ccc}
\ell_{1} & \ell_{2} & \ell_{3} \\
m_{1} & m_{2} & m_{3}
\end{array}\right),
\end{align}
where $(...)$ denotes the $3j$ symbol.

\bibliographystyle{refs}
\bibliography{references}

\end{document}